\documentclass[lettersize,journal]{IEEEtran}
\usepackage{amsmath,amsfonts}
\usepackage{algorithmic}
\usepackage{algorithm}
\usepackage{array}
\usepackage[caption=false,font=normalsize,labelfont=sf,textfont=sf]{subfig}
\usepackage{textcomp}
\usepackage{stfloats}
\usepackage{url}
\usepackage{verbatim}
\usepackage{graphicx}

\usepackage{caption}

\usepackage[
    backend=biber,
    style=ieee,
  ]{biblatex}

\usepackage{pifont} 

\newcommand{\xmark}{\ding{55}}

\begin{document}

\title{Physical Layer Security in Satellite Communication: State-of-the-art and Open Problems}

\author{Nora Abdelsalam, Saif Al-Kuwari, Aiman Erbad \\ 
College of Science and Engineering, Hamad Bin Khalifa University, Education City, Doha, Qatar
}

\markboth{}%
{Shell \MakeLowercase{\textit{et al.}}: Physical Layer Security in Satellite Networks: A Survey}

\maketitle

\begin{abstract}
Satellite communications emerged as a promising extension to terrestrial networks in future 6G network research due to their extensive coverage in remote areas and ability to support the increasing traffic rate and heterogeneous networks. Like other wireless communication technologies, satellite signals are transmitted in a shared medium, making them vulnerable to attacks, such as eavesdropping, jamming, and spoofing. A good candidate to overcome these issues is physical layer security (PLS), which utilizes physical layer characteristics to provide security, especially due to its suitability for resource-limited devices such as satellites and IoT devices. 

In this paper, we provide a thorough and up-to-date review of PLS solutions for securing satellite communication. We classify main satellite applications into five domains, namely: Satellite-terrestrial, satellite-based IoT, Satellite navigation systems, FSO-based, and inter-satellite. In each domain, we discuss and investigate how PLS can be used to improve the system's overall security, preserve some desirable security properties and resist popular attacks. 
Finally, we highlight a few gaps in the related literature and discuss open research problems and opportunities for leveraging PLS in satellite communication.

\end{abstract}


\section{Introduction}
\IEEEPARstart{A}{s} more devices are inter-connected, and the demand for high bandwidth and reliable communication is increasing. The current terrestrial networks have limited resources and spectrum, which  cannot meet the requirements of next-generation applications and the increasing number of internet of things (IoT) devices. In addition, current cellular networks cannot cover all geographical areas, especially remote locations with low populations, because network installation in these areas is both expensive and challenging \cite{wang2019convergence}.  As a result, satellite networks have gained tremendous interest in recent years as a promising solution for the limitations of terrestrial networks. 

Satellite communication systems cover large areas due to their broadcast nature, providing a large-scale footprint service and flexible connectivity \cite{wang2019convergence,li2019physical}. However, satellite links suffer from a significant round trip time that increases latency and decreases service quality. Therefore, using satellite links alone will not fulfill the expected network requirements. Leveraging the benefits of terrestrial and satellite networks, the researchers start designing new network architecture to achieve the quality of service required with ubiquitous connectivity. Furthermore, there is a significant interest in satellite constellation projects such as Starlink, which SpaceX initiated in 2015 to provide high-speed internet connection worldwide using LEO satellites\cite{spacex}. 



Signals sent by satellites are transmitted in a shared medium and can be received by all surrounding nodes, both legitimate and malicious. Hence, satellite communications are vulnerable to both passive and active attacks. In passive attacks, such as eavesdropping, the adversary observes the data transmitted through the channel, attempts to retrieve confidential information, and causes illegal leakage \cite{cao2020analysis}. On the other hand, in active attacks, such as spoofing, jamming, and data manipulation, the adversary tampers with the traffic during its transmission. For example, the attacker can pretend to be a legitimate satellite or station and obtain illegal access to sensitive information. These attacks can cause ripple effects to the satellite networks and enormous damage, such as distraction, denial of service, and economic losses.  
%


There are two approaches to secure such communication networks, the upper layer approach, and the physical layer approach. The upper layer approach is based on classical cryptographic techniques, including symmetric cryptography (using a secret key shared by legitimate users) and public key cryptography, and it proved stable against many widespread attacks. However, encryption and decryption are expensive processes requiring high computational power, which is not usually available for satellites and IoT devices \cite{shakiba2021physical}. Furthermore, key management and distribution are usually problematic for satellites because they require more complex architectures and protocols. Moreover, cryptographic techniques assume that the adversary does not have enough computational capabilities to break its schemes, such as RSA, making it vulnerable once fault-tolerant quantum computers are available \cite{wang2018survey}.   

On the other hand, physical layer security (PLS) techniques are based on the wiretap model introduced by Wyner \cite{wyner1975wire}, in which information-theoretic security can be achieved without pre-shared keys. This model has two channels: the main channel between the legitimate users (transmitter and receiver) and the wiretap channel between the legitimate transmitter and the eavesdropper. PLS aims to lower the quality of the wiretap channel compared to the main channel to maximize the mutual information difference between the two channels. PLS utilizes wireless channel characteristics such as reciprocity and spatial and temporal variations to encode the information exchanged.

\subsection{Existing Surveys}
PLS is a lightweight approach applicable for securing many wireless communication systems, such as satellites and IoT devices. Many works in the literature have recently focused on designing PLS schemes to secure terrestrial and satellite networks with different architectures. 

Some existing surveys mainly discuss satellite evolution from multiple perspectives. Surveys such as \cite{niephaus2016qos} and \cite{wang2019convergence} review the convergence of terrestrial and satellite network integration. The authors of \cite{niephaus2016qos} identify the critical building blocks and the applicability of the current networks for convergence from architecture and performance aspects. Similarly, \cite{wang2019convergence} is a comprehensive survey that discusses the motivation and requirements for the convergence of satellite and terrestrial networks and the key enablers. 

Another type of surveys focuses on the physical layer techniques and optimization approaches. The authors in \cite{liu2016physical} explain PLS fundamentals, including attack models and wiretap channel models. Similarly, the authors in \cite{hamamreh2018classifications} divide PLS types against passive attacks into the signal-to-interference-plus-noise ratio-based approach and complexity-based approach (key-based).
Likewise, the authors in \cite{wang2018survey} focus on the SNIR approaches and reviews works related to secure beamforming and antenna node selection. The authors in \cite{shakiba2021physical} provided a broader view of PLS techniques and reviewed the three main security properties: node authentication, message integrity, and confidentiality. 

Some surveys cover the security aspect of satellite communication \cite{li2019physical},\cite{tedeschi2021satellite},\cite{yue2022security}. Authors in \cite{yue2022security} discuss the vulnerabilities and security challenges of LEO satellite systems, including passive eavesdropping, active attacks, and interference with possible countermeasures. On the other hand, authors in \cite{tedeschi2021satellite} focus on the security enhancement techniques dividing it into physical layer security mechanisms and cryptography. They consider the PLS schemes that ensure confidentiality, authentication, and availability of satellite communications. Lastly, \cite{li2019physical} studies the state-of-art PLS implementations in three satellite network architectures: land mobile satellite networks, hybrid satellite and terrestrial relay networks, and integrated networks. A comparison of existing surveys is available in table 1. In our comparison, we focus specifically on PLS in satellite domains, which multiple surveys in the literature do not cover. 

\begin{table*}[!t]
\caption{Comparison between existing surveys and this survey \label{tab:table1}}
\centering
\begin{tabular}{|c|c|c|c|c|c|c|}
\hline
Survey & Year & IOT networks & GNSS satellites & satellite-terrestrial networks  & FSO satellites & inter-satellites \\
\hline
\cite{niephaus2016qos} & 2016 & \xmark & \xmark & \xmark & \xmark & \xmark \\
\hline
\cite{liu2016physical} & 2016 & \xmark & \xmark & \xmark & \xmark & \xmark \\
\hline
\cite{hamamreh2018classifications} &2018 & \checkmark& \xmark & \xmark & \xmark &\xmark \\
\hline
\cite{wang2018survey} & 2018 & \xmark & \xmark &\xmark &\xmark &\xmark \\
\hline
\cite{wang2019convergence} & 2019 & \xmark & \xmark& \xmark& \xmark & \xmark \\
\hline
\cite{li2019physical} & 2019 & \checkmark & \xmark & \checkmark& \xmark & \xmark\\
\hline
\cite{tedeschi2021satellite} & 2021 & \xmark & \checkmark & \checkmark & \xmark & \xmark\\
\hline
\cite{yue2022security} & 2022 &\xmark &\xmark & \xmark &\xmark &\xmark \\
\hline
Our survey & 2023 & \checkmark & \checkmark & \checkmark &  \checkmark & \checkmark \\
\hline
\end{tabular}
\end{table*}

\subsection{Contribution}

Extending terrestrial coverage is generally the main application of satellites. However, Satellites can help to realize many other important and practical applications. For example, satellite systems are vital in navigation applications, which continuously provide positions and time information for any location on earth. Global navigation satellite systems are a growing development field from the global positioning system (GPS) project in 1973 to regional systems projects planned to be launched in 2023 \cite{hein2020status}. Similarly, satellite-based IoT is a promising domain leveraging satellite coverage to connect internet of things devices distributed over large geographical locations and harsh environments, where  mission-critical IoT applications such as rescue operations or disaster recovery can finally be efficiently realized \cite{routray2019satellite}. 
Furthermore, satellites are prominent in optical communication, where free-space optical (FSO) networks are extended to cover increasingly larger areas via satellites. Eventually, satellite networks will further grow in size and maturity, and that will entail providing and maintaining communication between the satellites (inter-satellite communication), and that will inevitably lead to even more interesting applications.

Existing surveys provide reasonable coverage of the use of PLS at several satellite networks. However, the literature lacks a comprehensive survey covering multiple satellite-based domains, and this is what we attempt to address in this survey. 

In this paper, we classify satellite applications into five main application domains and analyze the PLS in each domain. These domains are: terrestrial and satellite networks, navigation satellite systems, satellite-based IoT, FSO-based systems, and inter-satellite communications. There are many other applications for satellites, such as those related to military and research, but to the best of our knowledge and based on our literature analysis, we found that these five areas are the main domains where satellites play a major role. Hence it is critical to analyze the security threats for each of these domains and study how PLS can provide the necessary protection, utilizing the unique characteristics of each domain.

The contributions of this survey can be summarized as follow.  

\begin{figure*}[!t]
\centering
\includegraphics[width=6in]{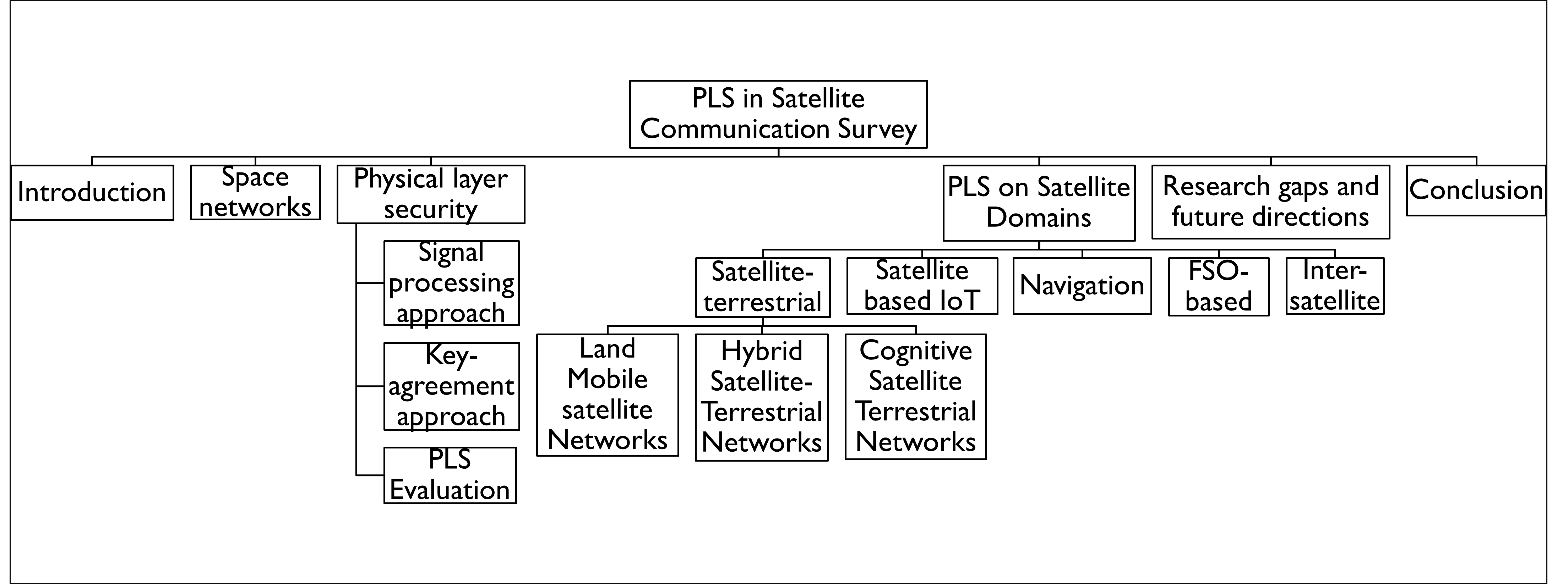}
\caption{Organization of this survey paper}
\label{fig_1}
\end{figure*}

\begin{itemize}
    \item We classify satellite networks as follows: Satellite-terrestrial, satellite-based IoT, Navigation, FSO-based, and Inter-satellite, as shown in Figure \ref{fig_1}. We highlight the primary usage for each domain and discuss their security concerns and vulnerabilities.
    \item We review the state-of-art works of PLS techniques for each satellite domain, considering their design and security analysis. We provide an intensive comparison between all available solutions for each area and discuss their goals and results.  
    \item We identify future trends and directions for PLS development in satellite networks. 
\end{itemize}

\subsection{Organization}
The rest of this paper is organized as follows: Section \ref{spacebackground} provides the necessary background for satellite and space networks while section  \ref{PLS background} provides background for PLS. In section \ref{mainsection}, we discuss the five domains of satellite networks. For each domain, we describe the architecture of the satellite network, its importance, and  review recent PLS schemes and compare them. Finally, we conclude this paper in section \ref{gaps} by highlighting open challenges and future directions of PLS in satellite networks. Figure \ref{fig_1} provides a visualization of the structure of this survey.

\section{Space Networks}
\label{spacebackground}
\begin{figure*}[!t]
\centering
\includegraphics[width=6in]{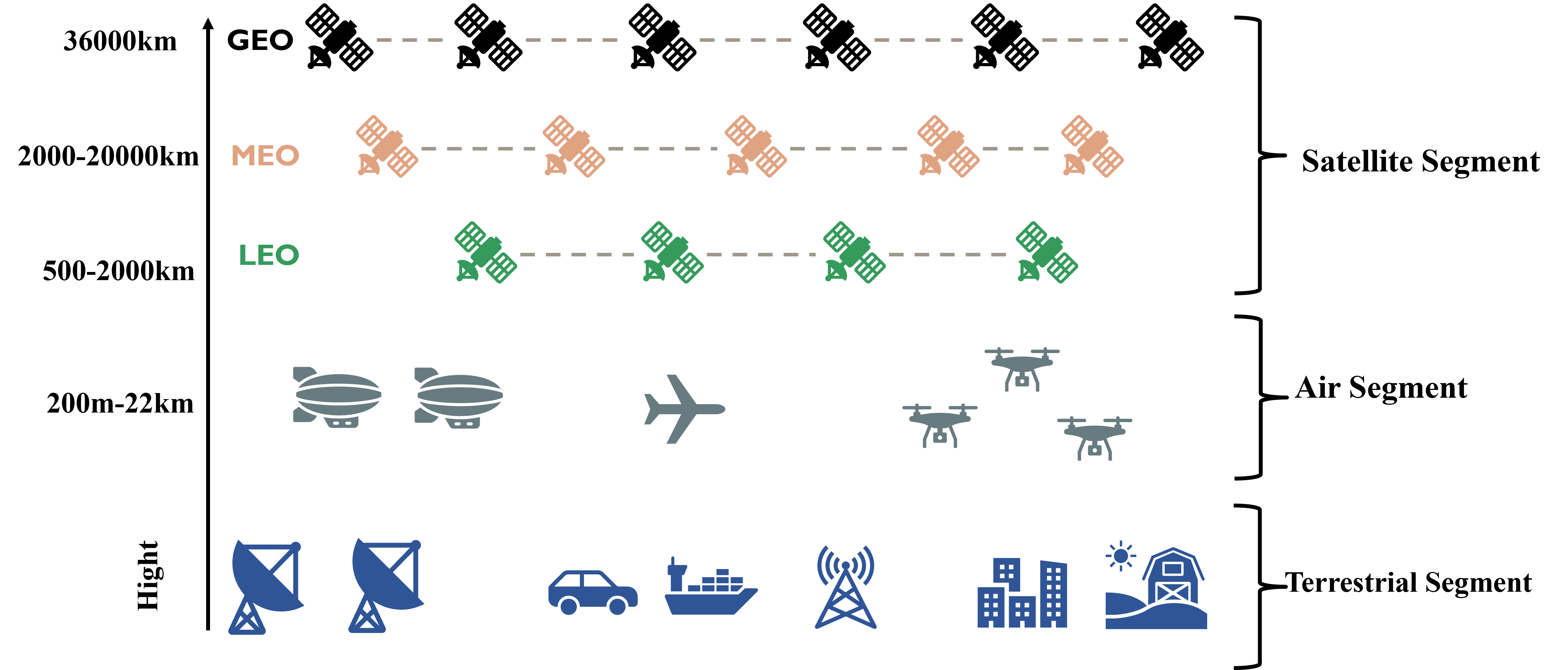}
\caption{Satellite and terrestrial integration architecture }
\label{fig_integ_arch}
\end{figure*}

Researchers propose several integration architectures to leverage the benefits of both satellites and terrestrial networks. A generic architecture illustrated in figure \ref{fig_integ_arch} extends the current system to achieve global coverage by dividing terrestrial-satellite networks into three main parts: terrestrial networks, air-based networks, and satellites network \cite{zhang2019multiple}, \cite{wang2019convergence} and \cite{bacco2019iot}. Terrestrial networks include conventional wired and wireless communication networks, satellite ground stations, and all user terminals on the earth's surface. The next level is the air-based networks, including High-altitude platform stations (HAPS), Unmanned Autonomous Vehicles (UAVs), and planes. The last level is the satellite networks, which include three types of satellites: low earth orbit (LEO), medium earth orbit (MEO), and geostationary earth orbit (GEO). 

Different satellite networks operate in different orbital heights resulting in different delays and coverage areas. GEO satellites operate at an orbit height of 36000 km, covering a large area of the earth, which increase their availability but introduces latency due to longer round trip. Hence, GEO satellites work as the space backbone network and are responsible for network management. MEO satellites operate in lower orbits at the height of 2000-20000 km, which decreases the latency, but increases the number of satellites needed to cover the earth. LEO satellites are the closest to the earth, operating on 500-2000 km, with decreased latency, path loss, and attenuation. Hence, LEO satellites are considered access points to space networks with inter-satellite links.

All these types of satellites with their communication links are vulnerable to several attacks. Like conventional adversaries, space adversaries can either be passive (attempting to compromise the confidentiality of the communication) or active (attempting to spoof the communication by tampering with the traffic). Furthermore,  adversaries can target the service availability of the satellite systems by jamming or compromising them. Therefore, various security concerns should be considered for each satellite network design and architecture. 

\section{Physical layer security (PLS)}
\label{PLS background}
The physical layer security concept was first introduced by Wyner \cite{wyner1975wire}, who proposed a wiretap channel model. This model secures legitimate users' channels from unauthorized eavesdropping or signal interception. It aims to utilize the wireless channel noise and randomness to outperform the quality of the eavesdropper's channel (wiretap channel). When the wiretap channel is degraded, the mutual information difference between the two channels is maximized, preventing the eavesdropper from being able to decode the transmitted signals. Mutual information is the amount of information the receiver obtains from the original message from the received one. Mutual information measures the reliability of the link because it quantifies the amount of information sent correctly to the receiver. Hence, PLS aims to maximize the mutual information of the legitimate link (increase reliability) while minimizing it on the wiretap link, which increases randomness and reduces the amount of information received by the eavesdropper \cite{guvenkaya2017physical}.  

PLS techniques are unconditionally secure (assuming a threat model where adversaries have access to unlimited resources), which makes them quantum-resistant by nature.  Furthermore, PLS does not require pre-shared keys between legitimate users, which solves the current key management and distribution difficulties. Finally, PLS is lightweight, making it an ideal choice for IoT applications.  

In the following section, we discuss some PLS approaches dominant in the literature. 

\subsection{Signal processing approach}
\begin{figure*}[!t]
\centering
\subfloat[]{\includegraphics[scale=0.3]{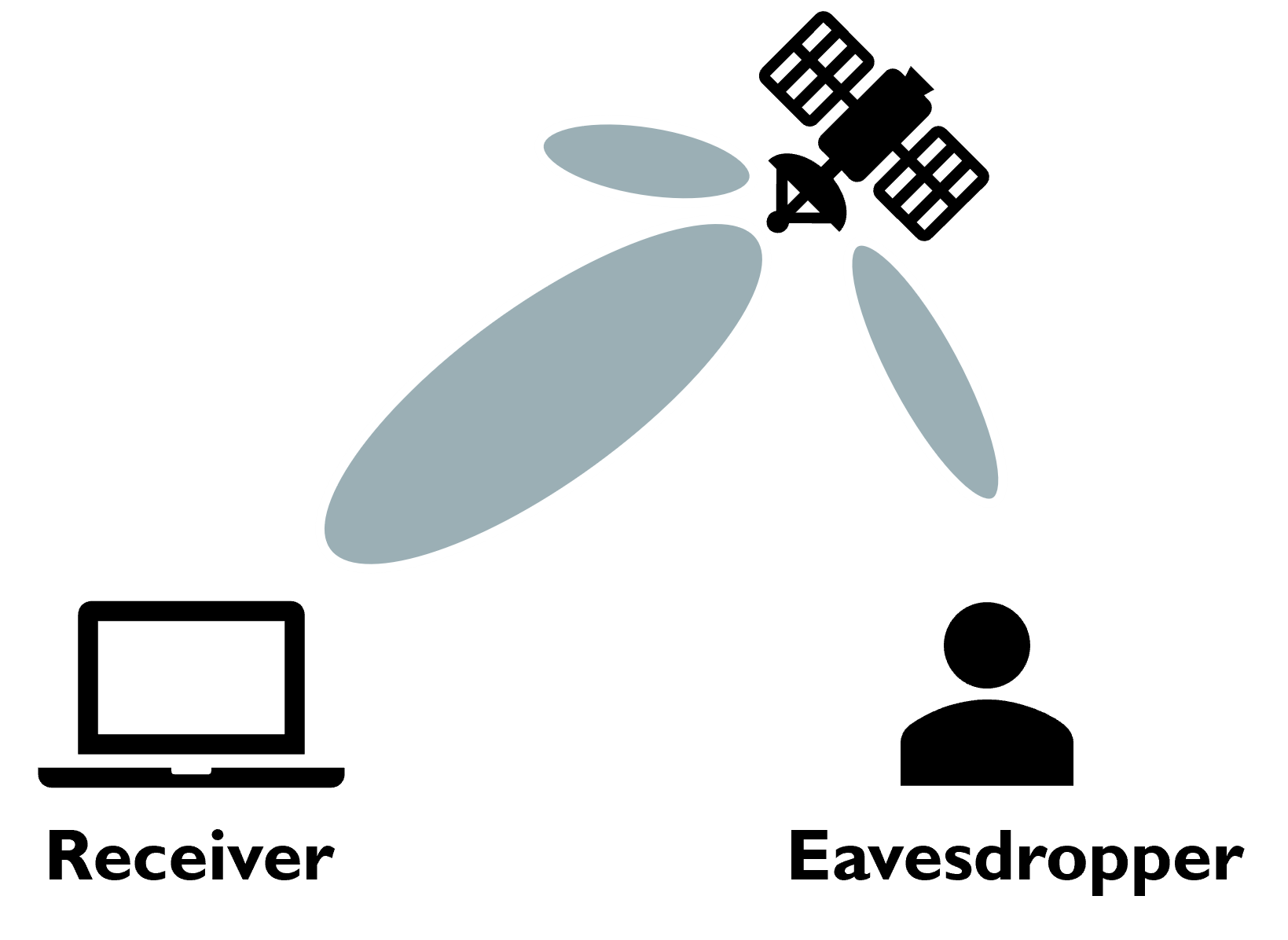}%
\label{beam}}
\hfil
\subfloat[]{\includegraphics[scale=0.3]{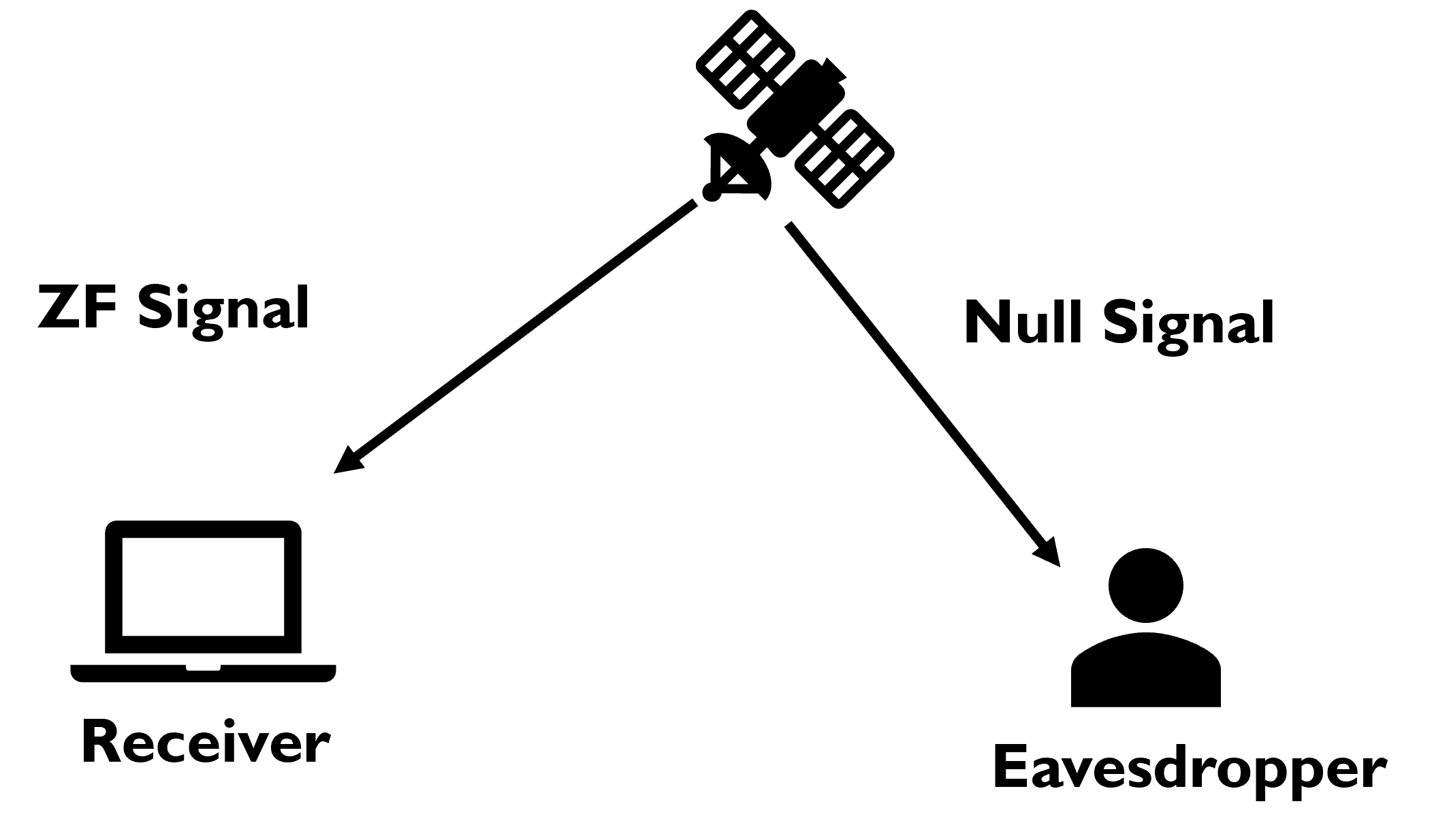}%
\label{ZF}}
\hfil
\subfloat[]{\includegraphics[scale=0.3]{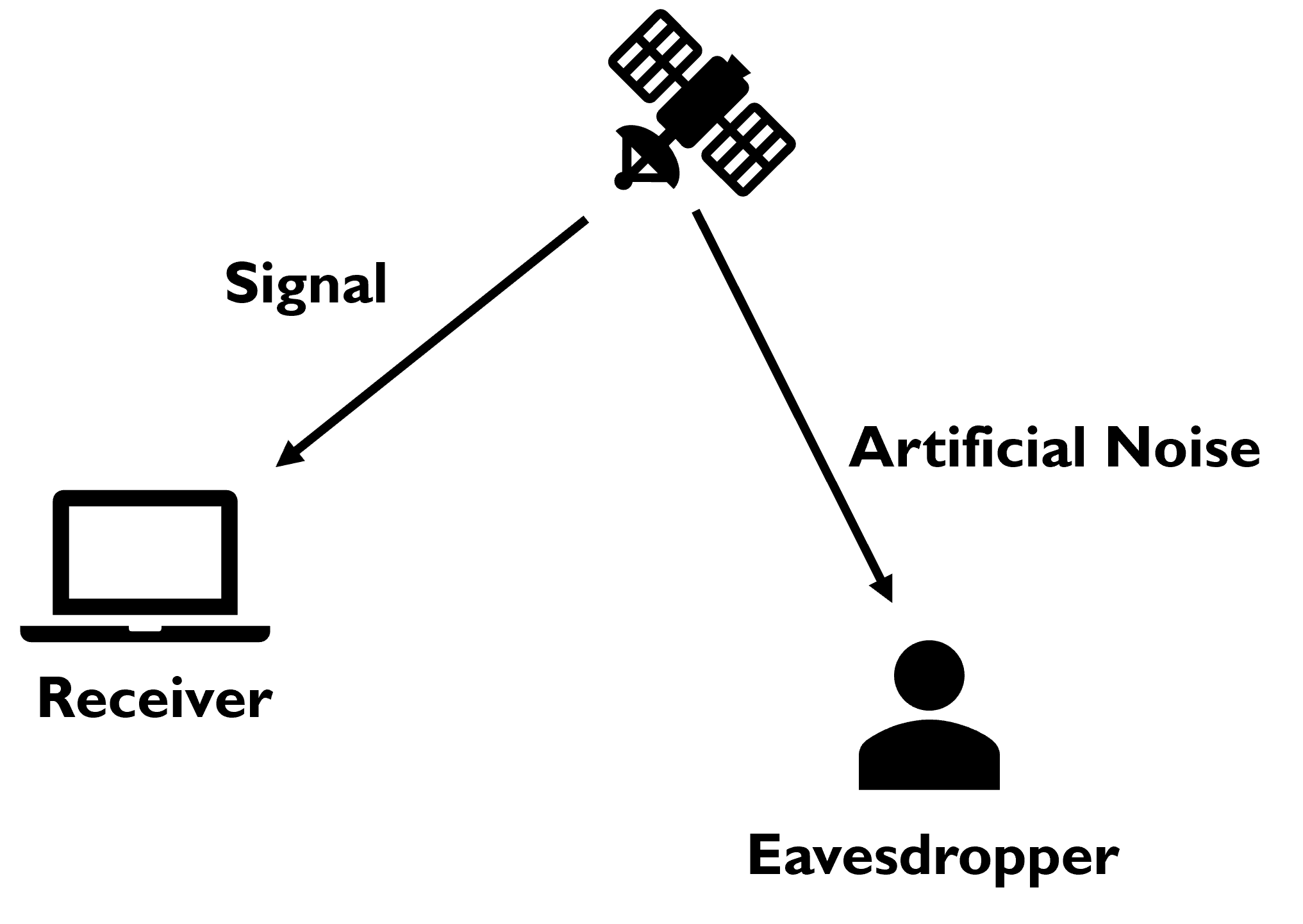}%
\label{AN}}
\caption{ Techniques for PLS based on signal processing (a) Beamforming technique that directs the signal towards the legitimate user (b) Zero-forcing technique sends messages to cancel interference orthogonal to the eavesdropper's (c) Artificial noise technique sends interference signal to the eavesdropper side.}
\label{techniques}
\end{figure*}

PLS signal processing approach increases the main channel's SINR (signal-to-interference and noise ratio)  compared to the wiretap channel. As a result, the eavesdropper will have a more noisy channel (degraded), which prevents her from accessing the transmitted data. This is a keyless approach and can be based on channel adaptation or inserting artificial signals to degrade the wiretap channel. 
An example of an enabling technology is multiantenna. 
%
Multiple input multiple-output (MIMO) is multiantenna technology where the transmitter and receiver have several antennas for communication. These antennas help in increasing the transmission rate and enhancing signal strength. Furthermore, MIMO techniques vary signal strength between legitimate users and eavesdroppers to increase secrecy capacity. MIMO technologies have recently attracted increasing attention in satellite communications \cite{li2021downlink} \cite{you2020massive}. Examples of PLS techniques based on MIMO are listed below:
 \begin{itemize}
 
\item{Beamforming:} a technique that focuses the signal strength in one direction creating a beam of signals using multiple antennas. This beam is used to direct the signal towards the legitimate user to maximize his SINR while downgrading the adversary's (as the latter only receives a negligible amount of leaked signal). Figure \ref{beam} illustrates how beamforming varies the signal strength between the two channels.
\item{Zero forcing:} a method for canceling the interference for multiple users in wireless communication. Zero forcing (ZF) precoding can be used to increase secrecy capacity by sending the message to cancel interference orthogonal to the eavesdropper's channel resulting in a null signal on the eavesdropper's side. Hence, this technique  creates the null signal based on the eavesdropper's channel state information (CSI). Figure \ref{ZF} illustrates how the ZF signal is orthogonal to the eavesdropper.
\item{Artificial noise:} an interference signal that increases the noise in the eavesdropper channel to protect transmitted data. The transmitter  will divide the transmission power for transmitting the message and sends the artificial noise in the direction of the eavesdropper. The transmitter needs to know the CSI of the eavesdropper to send this artificial noise. This noise is added orthogonal/ perpendicular to the data vector means in the null space of the receiver. Hence, it is designed to be canceled on the legitimate user side and only affects the illegitimate user. Figure \ref{AN} illustrates how artificial noise can be designed. 

\end{itemize}

Alternatively, the transmitter can use relay devices to amplify the signal on its way to the receiver \cite{bankey2021physical}. It can be used to increase the secrecy capacity through different cooperative strategies/communication with the transmitter. 

\subsection{Key-agreement approach}
key agreement techniques leverage the randomness of the channel between legitimate users to generate secret keys. In this method, there is no assumption that eavesdroppers will have a degraded channel or less SNIR. It works on systems where transmitter and receiver experience theoretically identical wireless channel states while it is different for eavesdroppers, which should be located half wavelength away \cite{hamamreh2018classifications}. In these systems, the users do not exchange explicit messages about the channel; they use only pilot messages, which the eavesdropper can use to estimate the channel and recover the key. This method utilizes channel state information (CSI), received signal strength (RSS), or phase shift between the legitimate user to extract randomness, which is then quantized into random bits. Next, a reconciliation process is used to remove erroneous bits. Finally, the generated key's randomness  is improved using privacy amplification, which removes any correlation between bits. This method is used to achieve confidentiality and authentication via the physical layer.

\subsection{PLS Evaluation}
Several security metrics for measuring the effectiveness of PLS techniques exist. In this section, we summarize the most commonly used security metrics with their definitions and expressions. 

\paragraph{Secrecy Capacity}
Secrecy Capacity is the highest communication rate that guarantees reliability for the legitimate link with perfect security, which means that mutual information for the eavesdropper channel is zero. Hence, it is the difference between the main channel capacity and the wiretap channel capacity. Secrecy Capacity can be expressed as:
    \begin{equation}
        C_s= max[I(X,Y_B) -I(X,Y_E)]
    \end{equation}
where $I$ is the mutual information, $X$ is the message sent, and $Y_B$ and $Y_E$ are the messages received by the legitimate user and the eavesdropper, respectively.

\paragraph{Achievable Secrecy Rate }
The achievable secrecy rate is the difference between the data rate achieved by the main channel and the wiretap channel with the Gaussian codebook. It can be expressed as: 
\begin{equation}
    R_s = [R_B- R_E]^+
\end{equation}{}

Where $[]^+$ means $max(x,0)$ and $R_B$, $R_E$ are the legitimate and eavesdropper link rates, respectively. The achievable secrecy rate is the lower bound of the secrecy capacity, which is used for more computationally affordable calculations since the difference in the secrecy capacity is a non-convex optimization problem \cite{wang2015enhancing}. 

\paragraph{Secrecy outage probability} 
Due to variations in channel state and quality, the secrecy rate can be affected over time. If the secrecy rate drops below a specific target rate, the secrecy outage happens, which indicates that the security goal can not be achieved. The secrecy outage probability can be measured as
\begin{equation}
    SOP=Pr(C_R <R_0)
\end{equation}{}
where $C_R$ is the channel capacity and $R_0$ is the target rate \cite{nguyen2015secrecy}. For some SOP is practical to know the likelihood of the secrecy outage to give more insights about the security level. 

\paragraph{Secrecy Energy Efficiency }
In power-constrained systems, it is essential to consider energy consumption in the security scheme design. Secrecy energy efficiency measures the energy consumption by all nodes with the security performance \cite{ouyang2017secrecy}. Specifically, it can be measured as
\begin{equation}
SEE= R_s / E_total 
\end{equation}{}

where $R_s$ is the secure bits, and E total is the total energy the system uses. Therefore, SEE measures the number of secure bits transmitted in the energy unit.  

\paragraph{Security Gap} 
The security gap is a practical measure of security performance that was introduced in \cite{klinc2009ldpc}. It is based on the bit error rate at the legitimate receiver and eavesdropper and can be expressed as: 
\begin{equation}
SG = SNR(B)_{min} - SNR(E)_{max}
\end{equation}{}

This equation denotes the difference between the minimum signal-to-noise ratio needed by the receiver to decode the signal correctly and the maximum signal-to-noise ratio for the eavesdropper that ensures a specific level of error rate. The gap between these two ratios indicates the quality advantage that legitimate users should have to guarantee the security requirement \cite{guvenkaya2017physical}. 

\section{PLS on Satellite Communication}
\label{mainsection}
In this survey, we classify satellite networks into 5 domains: Hybrid, IOT, Navigation, FSO, and inter-satellite. We review each of these domains in the next subsections. 
\subsection{Domain 1: Satellite-terrestrial Networks}
\label{sate_terr}
Extending terrestrial networks is the primary goal of satellite network integration to overcome current network limitations of coverage and spectrum. Specifically, leveraging the large footprint of satellites, the integration of satellite and terrestrial networks supports connecting rural and off-shores areas that current network infrastructure does not cover. Furthermore, the current congested spectrum poses serious challenges to applications requiring high bandwidth. Hence, spectrum sharing between satellite and terrestrial networks allows efficient spectrum use and provides reliable communication. 
However, several challenges must be solved to achieve this integration, including performance, resources, and security. Integration designs can be categorized into land mobile systems, hybrid and cognitive networks. In the following subsections, we will review each design architecture with its main security concerns and PLS works.  

\paragraph{Land Mobile satellite Networks }

\begin{figure*}[!t]
\centering
\subfloat[]{\includegraphics[width=3in]{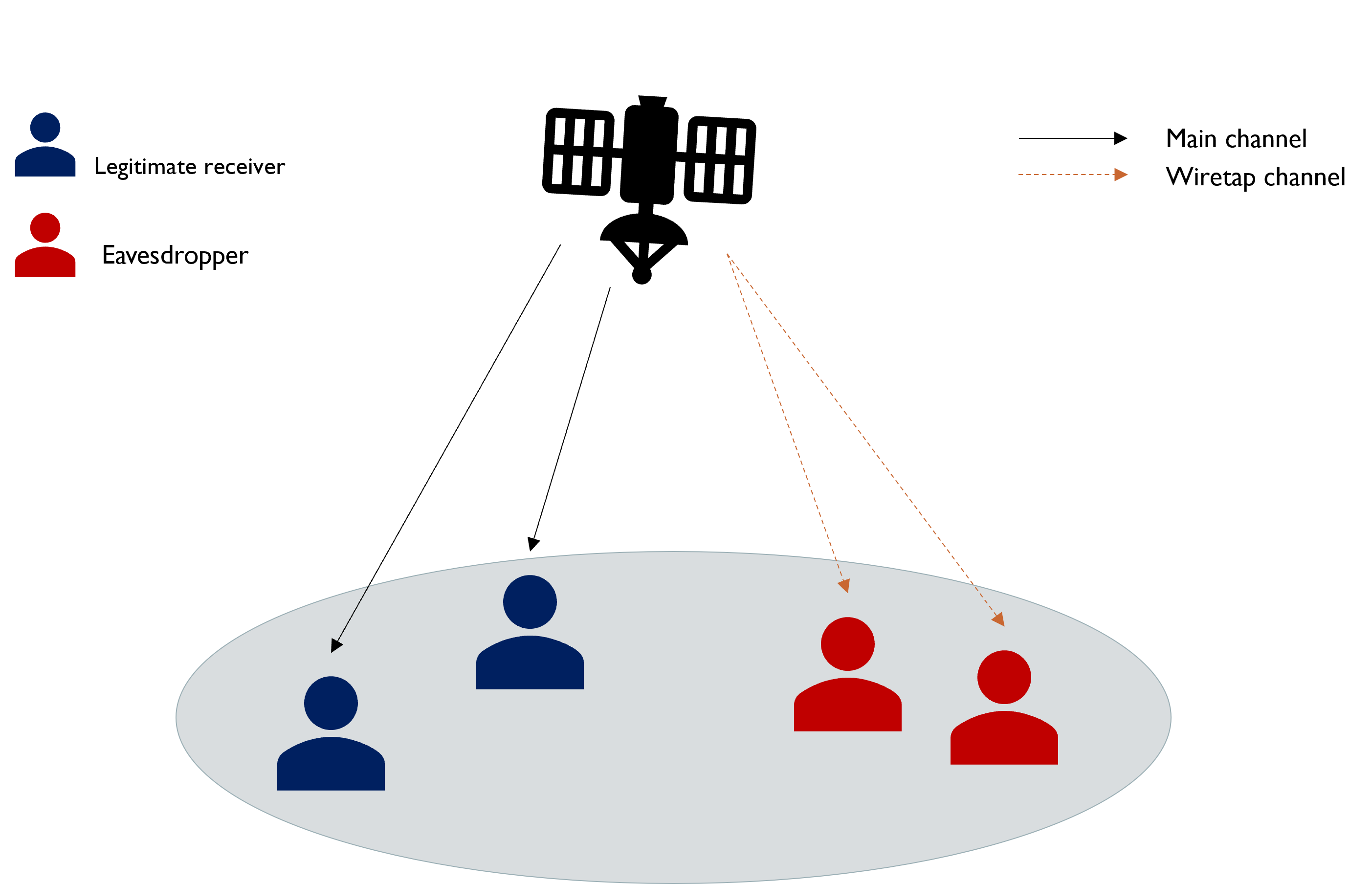}%
\label{fig_first_case}}
\hfil
\subfloat[]{\includegraphics[width=3in]{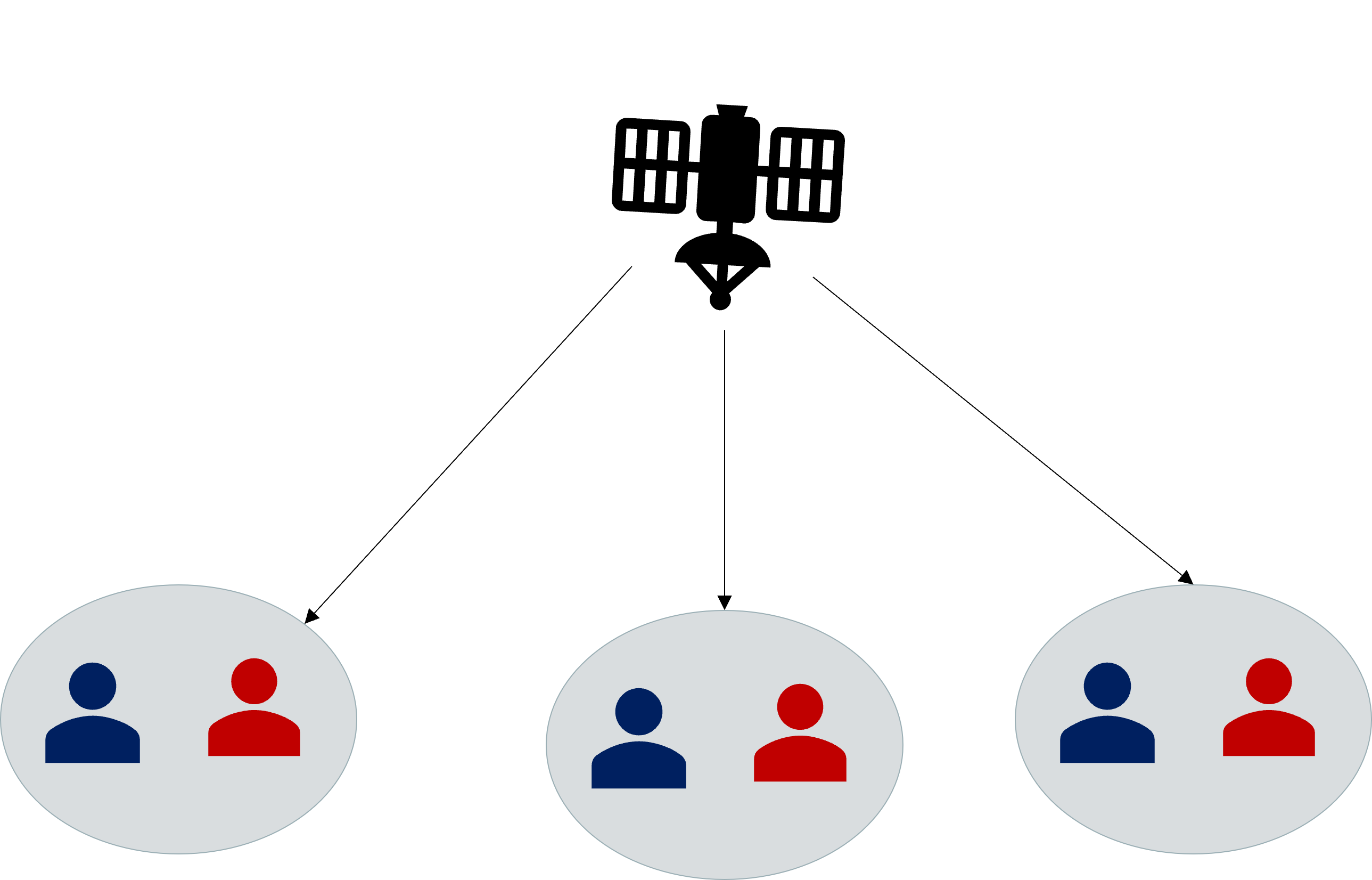}%
\label{fig_second_case}}
\caption{Land mobile satellite with multiple users and eavesdroppers (a) One spot beam satellite (b) Multi-beam satellite}
\label{landmobile}
\end{figure*}

Land mobile satellite (LMS) networks consist of satellites directly connected to ground stations using radio frequency signals. Satellites can be spot beams or multibeams, where the difference is the area of coverage and spectrum reuse. Specifically, multibeam allows coverage of multiple users using a single beam and sends the information to multiple spots on the ground with frequency reuse for each beam. Figure \ref{landmobile} shows the architecture of LMS with spot beam and multibeam satellites. As the figure shows, coverage of the satellite signals allows illegitimate users to receive the signal, which threatens link confidentiality and authenticity. Hence, several works focus on the security issues of LMS and study PLS  performance in such systems.

\cite{xiao2018secure} consider the security performance of Non-Geostationary satellites with different orbit models. Non-Geostationary satellites are not in a fixed position with respect to the earth, but it is moving around it in multiple orbits, such as LEO and MEO. They utilize satellite movement to evaluate the system's security in different positions and atmospheric conditions, including rain attenuation. They drive the closed-form expression of secrecy capacity and secrecy outage probability and provide Monte Carlo simulation.

Authors in \cite{vazquez2018physical} propose a wiretap channel model with a finite length regime. Their model includes a coding method that creates wiretap code using linear error-correcting codes such as polar codes and randomized hash functions. They consider RF satellite uplink and measure the secrecy capacity for the proposed scheme. Furthermore, their method identifies the spital regions where secrecy can be granted. Their analysis assumes the receiver satellite is on GEO or MEO orbits and the eavesdropper is in LEO, MEO orbits, or UAV in different heights.  

Subsequently, authors in \cite{wang2019physical} and \cite{guo2019physical} consider an LMS system with multiple users cooperating to receive the message, and multiple eavesdroppers are trying to intercept data sent to legitimate users. They obtain the closed-form expressions for the non-zero probability of secrecy capacity, the secrecy outage probability (SOP), and the average secrecy capacity (ASC). \cite{wang2019physical} extends the eavesdropping scenario to two scenes: colliding and non-colliding, where eavesdroppers collaborate together to wiretap the channel in the colliding case, and the best eavesdropper will be selected to wiretap the channel in the non-colliding scenario. Analysis in \cite{guo2019physical} considers imperfect channel estimation to study the effect of various fading scenarios on the secrecy performance. 

A recent attempt to leverage the advancements in MIMO technology in satellite PLS was conducted in \cite{schraml2020multiuser}, where they used a multibeam satellite with a single feed-per-beam (SFPB) antenna and full-frequency-reuse (FFR). This architecture allows the satellite to send multiple beams to the earth's surface with the same spectrum and polarization for all beams. They consider two designs for the antenna reflector: one parabolic reflector and multiple reflectors. They aim to maximize the secrecy capacity of each user against its eavesdropper, constrained by the power consumption for each beam. Hence they formulate the optimization problem and solve it using convex approximation. Furthermore, the authors introduce artificial noise to the multiple reflector case to increase secrecy when the number of eavesdroppers exceeds the number of beams. 

Giving more focus to authenticating satellite-to-ground links, authors in \cite{oligeri2020past} proposed a physical layer authentication mechanism to authenticate Iridium satellites from ground stations. Iridium is a constellation of 66 LEO satellites for voice and data transmission, and its signal can be received by dedicated mobile satellite devices or integrated transceivers. The proposed scheme depends on hardware impairments of the satellites making their IQ samples different and can be used as a fingerprint. They collect a dataset of IQ samples for each satellite in the formation and process it to create grayscale images. They authenticate satellites in two scenarios: 1) multiclass classification, where each satellite represents one class, and in this case, they use deep CNN for classification; 2) binary classification where they authenticate only one satellite against others, and here autoencoders are used. Their results showed that artificial intelligence algorithms can authenticate satellites based on their IQ samples. 

Similarly, \cite{fu2020initial} proposes a physical layer authentication method using the doppler frequency shift of the satellite. Specifically, their method focuses on authenticating the system information signaling (SIS) messages sent by satellite to allow the user equipment to request random access. Since illegitimate users also receive these messages, they can spoof the SIS messages by sending similar messages to the users, which will cause connection failure. The authentication scheme allows the user terminal to estimate the doppler shift of the received message (current channel) and compare it to the doppler shift calculated using satellite ephemeris and position. If both calculations are equivalent, then the satellite signal is authenticated. Their scheme outperforms the CSI-based authentication scheme due to the channel correlation drop.

Table \ref{tab:tableLMS} summarizes all discussed PLS works related to the LMS system. We compare all works against their security goal, satellite and link types, and their used PLS technique and measurement. The comparison shows that most works study confidentiality requirements more than other security goals and assume passive eavesdroppers that are not actively affecting the main channel. Moreover, all techniques focus on downlink security from satellite to ground without considering the uplink security despite their importance to the satellite networks.

\begin{table*}[!t]
\caption{Comparison between existing PLS works for land mobile satellite communication  \label{tab:tableLMS}}
\centering
\begin{tabular}{|m{0.7cm}|m{1.5cm}|m{2.1cm}|m{1.5cm}|m{2.2cm}|m{2.5cm}|m{3.5cm}|}
\hline
Work & Security Goal & Satellite Type & Link Type & PLS Techniques & PLS feature used  & Security Measurement \\
\hline
\cite{xiao2018secure} (2018) & Confidentiality & Non-geostationary orbit & Downlink & Information theory 
 & Orbiting models  &  \begin{tabular}{@{}c@{}}Security Capacity 
 \\ Secrecy Outage Probability\end{tabular}
 \\
\hline
\cite{vazquez2018physical} (2019) & Confidentiality  & RF satellite & Uplink & Wiretap coding & Eavesdropper’s noise power &  Security Capacity 
 \\
 \hline
 \cite{wang2019physical} (2019) & Confidentiality & Spot beam Single Antenna & Downlink & Information theory 
 & User cooperation &  \begin{tabular}{@{}c@{}c@{}}Security Capacity 
 \\ Secrecy Outage Probability \\ Average secrecy capacity\end{tabular}
 \\
\hline
 \cite{guo2019physical} (2019) &Confidentiality &Spot beam Single Antenna & Downlink & Information theory 
 & Channel Estimation errors
 &  \begin{tabular}{@{}c@{}c@{}}Security Capacity 
 \\ Secrecy Outage Probability \\ Average secrecy capacity\end{tabular}
 \\
\hline

\cite{oligeri2020past} (2020) & Authentication & LEO satellite & Downlink  & Hardware Fingerprinting & IQ samples & Classification accuracy 
 \\
\hline
\cite{fu2020initial} (2020) & Authentication &LEO satellite & Downlink & Physical layer authentication & Doppler frequency shift
 & False alarm rate miss detection rate(FAR) 
\\
\hline
\cite{schraml2020multiuser} (2021) & Confidentiality & Multibeam  & Downlink & MIMO precoding - Artificial noise & Anntena spatial degree of freedom& Minimum Secrecy capacity 
\\
\hline
\end{tabular}
\end{table*}

\begin{figure}[!t]
\centering
\includegraphics[scale=0.3]{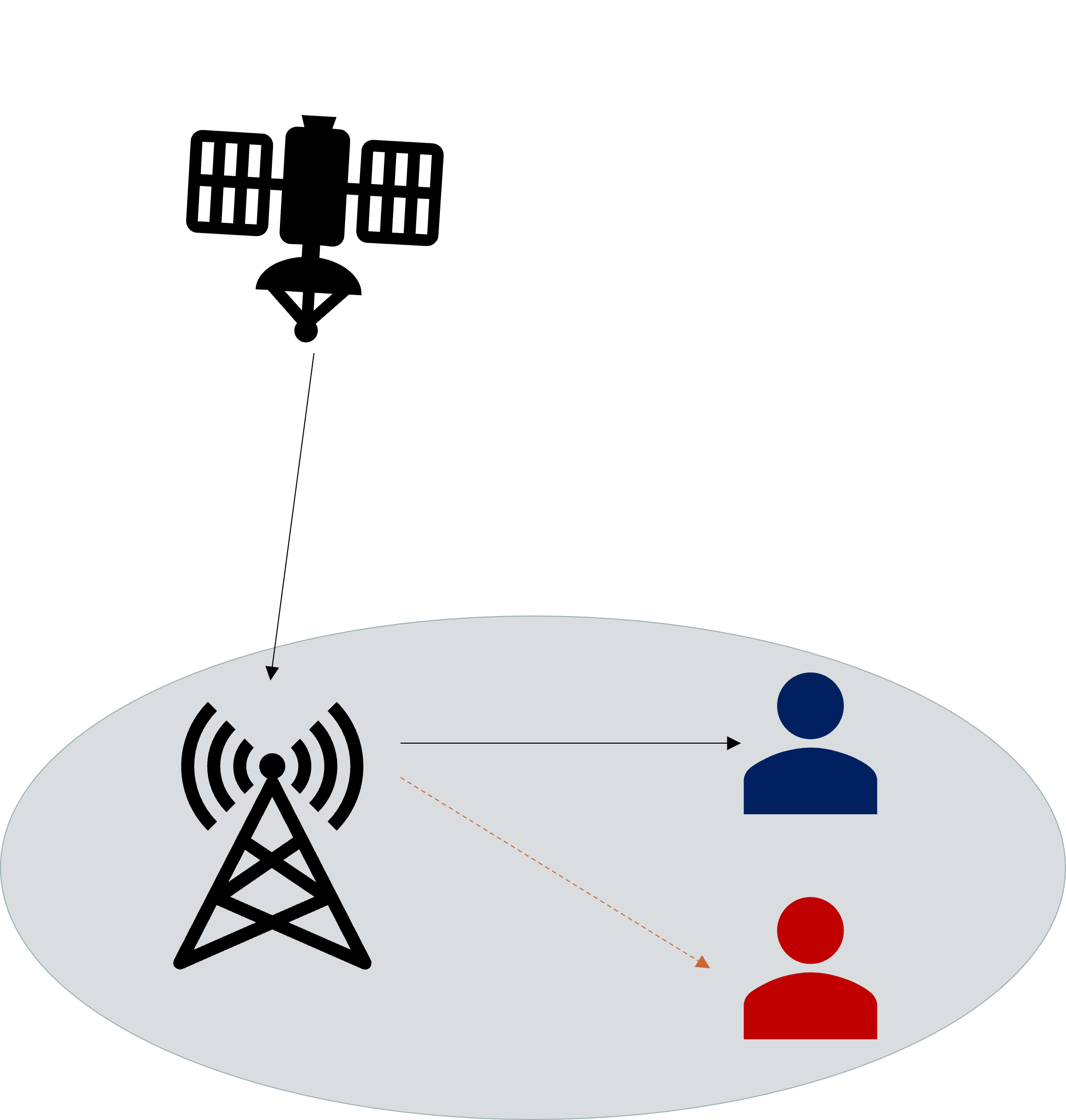}
\caption{Relay-based satellite-terrestrial system }
\label{fig_relay}
\end{figure}

\paragraph{Hybrid Satellite-Terrestrial Networks }
In hybrid satellite-terrestrial networks (HSTNs), the satellite transmits the data to the ground destination with the help of a middle relay, as shown in figure \ref{fig_relay}. This relay can be terrestrial stations or air space devices such as UAVs or HAPS. Relays can help to improve the security rates using several techniques.

\begin{itemize}
    \item Amplify and forward (AF): This method can be used when the transmitter has limited power to send the message. The relay cooperates with the transmitter by amplifying the transmitter's signal (message) and sending it to the receiver without decoding it. As a result, the transmission rate increases, and the receiver's noise and interference decrease.
    \item Decode and forward (DF): In this type, the relay cooperates with the transmitter by decoding the received message, then re-encodes it and sends it to the receiver. DF relay location affects the secrecy of the link; the closer the transmitter and the relay are, the higher the secrecy of the link \cite{lai2008relay}. 
    
    \item Noise and forward (NF): In this type, the relay has two channels. One sends the message to the receiver, and the other sends artificial noise to the eavesdropper's channel. In some cases, this type is used only to send AN to confuse eavesdroppers and increase secrecy capacity.
\end{itemize}
HSTN is considered a practical design for terrestrial and satellite integration, encouraging researchers to study its performance and security since it is vulnerable to several attacks. 

Authors in \cite{guo2018physical} proposed a relay selection and user scheduling joint scheme and studied its security capacity with multiple users, relays, and eavesdroppers. Their scheme depends on measuring the signal-to-noise ratio (SNR) on both links from the satellite to the relays and from the relay to the users in order to choose the best link. They assume that eavesdroppers can cooperate to access legitimate information in the first scenario. In contrast, in the second scenario, the eavesdropper with the best signal-to-noise ratio will wiretap the channel. Furthermore, the authors in \cite{cao2018relay} propose two models of relay selection: single relay and multi-relay, and compare them to the baseline round-robin scheduling. The single relay scheme chooses the best relay based on the link's instantaneous capacity between the relay and the destination since they assume a passive eavesdropper with unknown channel information. However, in the multi-relay scheme, multiple relays transmit the data simultaneously with a weighted version of the decoded message. Their schemes outperform the round-robin method. Similarly,\cite{huang2021reliability} investigates the security and reliability of two relay selection schemes in an HSTN system with artificial noise and compares them to the round-robin baseline under different shadowing conditions. They drive the closed-form expressions of the outage probability and intercept probability and present a numerical analysis of the two schemes. Their results showed that increasing the number of relays will increase the reliability of the two schemes while reducing the security where intercept probability increases. 

\cite{bankey2019physical} propose a PLS framework for downlink HSTN using two relay techniques: amplify and forward,or decode and forward with multiple relays and users. Here the satellite has multiple antennas where it transmits its signal to all relays, and only the selected relay will amplify or decode the signal and forward it to the user. They assume that eavesdroppers can only wiretap the communication between the relay and users (terrestrial link) with two scenarios of eavesdropping, colluding, and non-colluding. The relay selection mechanism depends on the signal-to-noise ratio between the user and eavesdropper. Their analysis showed that secrecy outage probability is worse in the colluding scenario due to the cooperation between eavesdroppers. Moreover, increasing the number of users enhances the security performance due to the increased diversity gain. 

Authors in \cite{li2019physical2} add power allocation as a constraint to the relay selection scheme to minimize secrecy outage probability in HSTN using instantaneous and statistical CSI. In this paper, the relay used is near ground relays, such as airplanes or HAPS, and it transmits an interference signal to downgrade the channel for the eavesdropper with no effect on the quality of the legitimate user channel. Since instantaneous CSI can not be obtained in satellite communication, they propose a relay selection scheme based on statistical CSI to decrease the signal overhead and delay caused by continuous relay handover. Besides, statistical CSI is used for satellite and relay communications power allocation to minimize the system's energy consumption while enhancing the secrecy level. Similarly, authors in \cite{cao2022secrecy} use the known and unknown CSI conditions for the wiretap channel to propose two relay-user pairing schemes. The optimal relay user pairing scheme assumes complete knowledge of eavesdropper channel state information, hence the relay that can maximize the secrecy rate will be selected. However, the suboptimal method assumes that the full CSI of the eavesdropper is unknown since its a passive receiver. Hence it depends only on the main's channel information to choose the best relay.  

Authors in \cite{wu2019secrecy} consider a different perspective where they study the effect of the hardware impairments on the secrecy performance of an HSTN. They propose an optimal relay selection scheme specifically for the case where the hardware of nodes is not ideal. Their results show the superiority of the proposed scheme over the traditional round-robin scheduling for both ideal and impaired hardware cases. Moreover, they showed that in a high SNR situation, the secrecy capacity becomes lower in case of impaired hardware, increasing the secrecy outage probability. However, by increasing the number of relays, the secrecy performance enhances. In \cite{bankey2020physical}, the authors study the secrecy performance of NOMA-based users in the HSTN system. They consider the secrecy outage probability and define it as when a user fails to achieve the required level of secrecy. Their results showed that the channel condition of the week user does not affect the secrecy performance. 

Utilizing a different type of relay, \cite{sharma2020secure} proposed a 3D UAV relay HSTN system with an aerial eavesdropper around the UAV. The eavesdropper keeps track of the UAV relay to place himself at a suitable distance to get the transmitted data. They consider two scenarios where the eavesdropper is in a fixed place and a random place. They propose three strategies for UAV relay selection. The optimal method is based on the maximum-signal-to-noise ratio for the destination since the eavesdropper CSI is assumed to be unknown. However, they propose two less complex selection schemes because of the delay in channel estimation in satellite communications. One approach is based on the distance between the UAV relay and the destination and selecting the closest relay for transmission. The other approach chooses the relay randomly to assist the communication. Their results showed that selection based on SNR has the best performance and the closest distance-based selection method outperforms the random selection. Furthermore, they found that the secrecy performance of the mobile UAV relaying outperforms the static conventional relaying. All discussed works assume no direct link between the satellite and the users due to shadowing and masking effects. 

Table \ref{tab:tableHybrid} summarizes discussed work for PLS related to hybrid satellite-terrestrial networks. We compare the adversary model of all works and show that they all consider passive adversary models using single or multiple eavesdroppers with a single antenna. Furthermore, all  works use the wiretap channel's signal-to-noise ratio and channel state information to ensure security. However, these features can be challenging to measure in the case of hidden adversaries. Similar to the previous section, all  works focus on the confidentiality requirements only without studying other security aspects, such as authentication of the relay or satellite. 

\begin{table*}[!t]
\caption{Comparison between existing PLS works for Hybrid satellite-terrestrial communication  \label{tab:tableHybrid}}
\centering
\begin{tabular}{|m{0.7cm}|m{1.6cm}|m{1.9cm}|m{3.2cm}|m{1.9cm}|m{2.5cm}|m{3.7cm}|}
\hline
Work & Security Goal & Relay Type & Adversary Model & PLS Techniques & PLS feature used  & Security Measurement \\
\hline
\cite{guo2018physical} (2018) &Confidentiality& \begin{tabular}{@{}c@{}} Ground based\\ DF \end{tabular} & \begin{tabular}{@{}c@{}c} Multiple eavesdropper \\ Colluding and Non-colluding\end{tabular}& Relay technology& Signal to Noise ratio&Security Capacity 
 \\
\hline

\cite{cao2018relay} (2018) & Confidentiality& \begin{tabular}{@{}c@{}} Ground based\\ DF \end{tabular} & \begin{tabular}{@{}c@{}} Single passive eavesdropper
\\ Wiretap both links \end{tabular}& Relay technology & Channel state information&Secrecy outage
 probability
\\
\hline
\cite{bankey2019physical} (2019) & Confidentiality& \begin{tabular}{@{}c@{}} Ground based\\ DF, AF \end{tabular} & \begin{tabular}{@{}c@{}} Multiple eavesdropper \\ Colluding and Non-colluding\end{tabular}& Relay technology - Multi-antenna & Signal to Noise ratio & Secrecy outage
probability 
\\
\hline

\cite{wu2019secrecy} (2019) & Confidentiality& \begin{tabular}{@{}c@{}} Ground based\\DF \end{tabular} & \begin{tabular}{@{}c@{}}Single antenna eavesdropper\\Wiretap relay link\end{tabular}& Relay technology& Signal to Noise ratio & Secrecy outage
probability
\\
\hline

\cite{bankey2020physical} (2020) & Confidentiality& \begin{tabular}{@{}c@{}} Ground based\\ AF \end{tabular} & \begin{tabular}{@{}c@{}} Single antenna eavesdropper\\ Wiretap relay link \end{tabular}& Relay technology& Signal to Noise ratio &\begin{tabular}{@{}c@{}c@{}}Secrecy outage probability\\ Positive secrecy capacity \\probability  \end{tabular}
\\
\hline
\cite{li2019physical2} (2020) &  Confidentiality& \begin{tabular}{@{}c@{}} Aerial\\Interference relay\end{tabular} & \begin{tabular}{@{}c@{}} Near user eavesdropper\\ Wiretap Satellite link \end{tabular}& Relay technology -
 Artificial Noise& Channel state information (instantaneous  statistical) &Secrecy outage probability
\\
\hline
\cite{sharma2020secure} (2020) & Confidentiality& \begin{tabular}{@{}c@{}} 3D mobile UAV \\ DF\end{tabular} & \begin{tabular}{@{}c@{}} Aerial eavesdropper in fixed \\  \& random positions \end{tabular}&  Relay technology& \begin{tabular}{@{}c@{}c@{}}Signal to Noise ratio\\ UAV to user distance\end{tabular} &\begin{tabular}{@{}c@{}c@{}} Secrecy outage probability\\ Non-zero secrecy capacity\\ probability \end{tabular}
\\
\hline
\cite{huang2021reliability} (2021) & Reliability and confidentiality & \begin{tabular}{@{}c@{}} Ground based \\ DF\end{tabular} & \begin{tabular}{@{}c@{}} Single eavesdropper \\ Wiretap Satellite link  \end{tabular}&  Relay technology - Artificial Noise& Signal signal to interference plus-noise ratio &\begin{tabular}{@{}c@{}} Outage probability\\Intercept probability \end{tabular}
\\
\hline
\cite{cao2022secrecy} (2022) &  Confidentiality& \begin{tabular}{@{}c@{}} Ground based \\ DF\end{tabular} & \begin{tabular}{@{}c@{}} Single eavesdropper \\ Wiretap Both link  \end{tabular}& Relay technology& Channel state information (known \& unknown) & Secrecy outage probability
\\
\hline
\end{tabular}
\end{table*}

\paragraph{Cognitive Satellite Terrestrial Networks}

 Increased demand for bandwidth due to new applications and the advancement of communication systems cause spectrum congestion in terrestrial and satellite networks. Currently, each of these networks is working on their respective spectrums, and congestion is challenging for both networks. Cognitive satellite-terrestrial networks (CSTN) are based on the spectrum sharing between the two networks to solve the spectrum congestion issue by effectively using current frequency bands \cite{kolawole2017performance}. Cognitive radio is a promising paradigm and uses three primary schemes: underlay, overlay, and interweave. For satellite-terrestrial integration, underlay is widely considered to keep the interference of the terrestrial network to the satellite user under an acceptable level \cite{an2016secure}. Although interference is a challenge in the cognitive network, it can be beneficial to enhance secure transmission if appropriately designed to degrade the wiretap channel more than the main channel. In this section, we review the current state of art PLS works in CSTN. 
 
 Authors in \cite{lin2018joint} proposed two beamforming techniques that utilize the interference from the terrestrial network to improve the PLS security of the satellite network with joint optimization of the transmit power and the quality of service. Their network architecture is software-defined based cognitive satellite-terrestrial where a gateway is the control center to manage the network with a database server that handles the system's CSI and spectral data. They consider the single and multiple eavesdropper scenarios and use the penalty function approach to achieve the required beamforming weight vectors for the base station according to the data in the database server. Authors in \cite{nguyen2020reliable} consider NOMA-based CSTN with decode and forward relay as a secondary network with hardware impairments and a passive eavesdropper. Specifically, they drive the closed-form expression of outage and inception probability with multiple antenna designs.

Both works \cite{li2018robust} and  \cite{yan2020robust} propose a robust beamforming framework for CSTN with probabilistic quality of service constraints related to the user, eavesdropper, interference limit, and transmission power consumption. Both works use a terrestrial base station for beamforming to enhance the PLS. However, \cite{li2018robust} used the terrestrial network as a source of friendly interference to secure the satellite user link in the presence of an eavesdropper. While in \cite{yan2020robust}, the eavesdroppers wiretap terrestrial links instead of satellite links in simultaneous wireless information and power transfer (SWIPT) enabled network. The eavesdroppers are passive in both cases, with imperfect channel state information and angle based in the second case. The primary goal of the beamforming technique in \cite{li2018robust} is to minimize the transmit power while achieving at least minimum SINR to the satellite user, maximize the tolerance to SINR leakage to the eavesdropper, and constrain the interference power to the satellite user. Similarly, \cite{yan2020robust} paid more attention to maximizing the achievable secrecy rate in the worst case for all terrestrial users while achieving the required SINR and Energy harvesting requirements for each user. Besides, they add a constraint on base station power consumption and interference limit for satellite stations. Due to these constraints, the problem becomes non-convex, so the authors use mathematical models to approximate the problem and then evaluate the computational complexity of their scheme and show its validity through numerical results.

Different from the above techniques, \cite{lin2020secure} add the assumption of an imperfect channel state for users and earth stations, not only the eavesdroppers, and propose a beamforming technique to exploit the base station as a source of green interference. Specifically, they designed a cooperative jamming scheme using a base station and cooperative terminal to degrade possible wiretap channels in the beamforming region, constraining the total power required for both terminals and SNIR and the interference limit for legitimate users. \cite{bouabdellah2021phy} use a jamming-based technique to enhance the security of CSTN where a satellite receives the signal from a terrestrial user through RF links and then forwards it to another ground station using an optical link in the presence of an active eavesdropper for each link. They assume that one base station acts as a friendly jammer that uses pseudo-random sequences known only to legitimate users to generate artificial noise that affects illegitimate inception since they cannot decode it. The authors drive the closed form and asymptotic expressions of inception probability considering the presence and absence of the jammer. 

Authors in \cite{lin2020robust} and \cite{lin2021secrecy} consider imperfect departure angles for the wiretap channel in a multibeam satellite system that shares a portion of the MMwave band with the cellular network to propose a secure beamforming scheme. Since beamforming techniques affect the system's power consumption, both works consider energy optimization in their design. In \cite{lin2020robust}, they consider a wireless information-powered network (WIPT) and focus on the power minimization problem with constraints on SINR level for users and secrecy limit to energy receivers as potential eavesdroppers. However, in \cite{lin2021secrecy}, they use a hybrid digital and analog beamforming design to achieve a cost-performance trade-off and aim to maximize secrecy energy efficiency, which is the ratio between secrecy rate and consumed power. Both works use a cloud processing center for resource allocation and control of the integrated network CSTN. Sequential convex approximation (SCA) was adopted to convert the non-convex problems into convex ones, and the results showed the advantage of the proposed schemes in different scenarios. 

Table \ref{tab:tableCognitive} summarize the discussed PLS works related to CTSN. 
Similar to other satellite-terrestrial networks, the cognitive network works focus on the system's confidentiality and study the cases of single and multiple eavesdroppers. However, the energy efficiency constraint is vital for cognitive networks since most works use beamforming and artificial noise techniques that require additional energy and can affect system performance. 

\begin{table*}[!t]
\caption{Comparison between existing PLS works for Cognitive satellite-terrestrial communication  \label{tab:tableCognitive}}
\centering
\begin{tabular}{|m{0.7cm}|m{1.6cm}|m{5em}|m{3.2cm}|m{2.2cm}|m{2.4cm}|m{3.3cm}|}
\hline
Work & Security Goal & Network Type & Adversary Model & PLS Techniques & PLS feature used  & Security Measurement \\
\hline
\cite{lin2018joint} (2018) & Confidentiality& Software Defined Network & Single and Multiple eavesdropper & Beamforming & Channel state information & Secrecy rate
 \\
\hline

\cite{li2018robust} (2018) & Confidentiality& Downlink CSTN & \begin{tabular}{@{}c@{}} Single antenna eavesdropper
\\ Wiretap satellite link \end{tabular}& Beamforming &  Probabilistic channel state information & SNIR constraints
on user and eavesdropper
\\
\hline
\cite{nguyen2020reliable} (2020) & Confidentiality & NOMA based CSTN& \begin{tabular}{@{}c@{}} Multi-antenna eavesdropper \\ Wiretap terrestrial link \end{tabular}& Relay technology and Multi-antenna & Hardware impairments 
&\begin{tabular}{@{}c@{}c@{}} Secrecy outage probability\\ Intercept probability \end{tabular} 
\\
\hline

\cite{yan2020robust} (2020) &Confidentiality&  mmWave CSTN with  SWIPT & \begin{tabular}{@{}c@{}} Multiple eavesdroppers\\ Wiretap terrestrial link\end{tabular}& Beamforming&Angle-based imperfect channel state information
 & Achievable secrecy rate
\\
\hline

\cite{lin2020secure} (2021) & Confidentiality& mmWave CSTN
 & \begin{tabular}{@{}c@{}} Multiple eavesdroppers \\Wiretap terrestrial link \end{tabular}&Beamforming and Artificial Noise& Imperfect channel state information &Signal to noise ratio
\\
\hline
\cite{bouabdellah2021phy} (2021) & Confidentiality & Underlay CSTN & \begin{tabular}{@{}c@{}} Two eavesdroppers \\ Wiretap satellite links \end{tabular}&  Artificial Noise& Pseudo-random sequences & Inception probability
\\
\hline
\cite{lin2020robust} (2021) & Confidentiality & WIPT Cloud processing architecture & Energy receivers & Beamforming & Angle-of-departure based channel state information &Achievable secrecy   rate
\\
\hline
\cite{lin2021secrecy} (2021) & Confidentiality& mmWave CTSN  Cloud processing architecture & \begin{tabular}{@{}c@{}} Multiple eavesdropper \\ Wiretap terrestrial link\end{tabular}& Hybrid Beamforming &Angle-of-departure based channel state information & Secrecy energy efficiency
\\
\hline
\end{tabular}
\end{table*}

\subsection{Domain 2: Satellite-based IoT Networks}
\label{Iot}
IoT communications are expected to provide massive connectivity in several services, including transportation, smart homes, manufacturing, health, agriculture, and maritime. These applications require extensive coverage in remote areas that terrestrial networks cannot support. In this case, the satellite network is a promising solution to provide IoT systems with broad coverage, including in harsh and rural areas. Specifically, LEO satellites can provide a trade-off between coverage and round trip latency around the earth, making them more favorable than MEO and GEO\cite{centenaro2021survey}. Moreover, hybrid satellite and terrestrial networks are proposed for maritime IoT to increase transmission efficiency and provide maritime service \cite{wei2021hybrid}. 
However, several challenges exist to achieve this connection, such as physical layer design and medium access mechanism. Security attacks are a severe concern in satellite IoT systems where IoT devices are considered vulnerable points exposed to different types of attacks such as interruption, spoofing, eavesdropping, and denial of service attacks \cite{lu2018internet}. For example, IoT devices can be used to maliciously overwhelm one satellite by sending lots of packets, causing a denial of service attack and interruption in the service. Applying intensive cryptographic techniques to satellite IoT devices is inapplicable because of the power limitation of these devices and the short characteristics of their packets. In the following section, we discuss how physical layer security is used to secure satellite-IoT communication through recent works. 

In \cite{yin2019secrecy}, the authors studied the physical layer secrecy analysis for satellite downlink signals with multiple mobile users and eavesdroppers in areas not covered by terrestrial networks. They proposed the FD-NOMA scheme that uses frequency division of downlink spectrum to provide efficient and secure multiple access for multi-users using partial spectrum sharing between users. This scheme increases the inter-user interference, which is leveraged in PLS to downgrade the eavesdropper's channel, while they propose a cooperative scheme that preserves the signal-to-noise ratio for legitimate users by interference cancellation. They drive the closed form of SINR for legitimate users and define a lower bound of secrecy rate for the system. In \cite{bas2019physical}, they proposed a similar scheme to introduce artificial interference and leverage it to affect the eavesdropper channel, using overlapping pulse shapes. They analyze the mutual information and secrecy capacity if the eavesdropper cannot resolve the time packing interference and in case he has an estimation module, and in both cases, security capacity is achieved.

\cite{ruan2020cooperative} consider cognitive satellite-terrestrial network for IoT services with a UAV malicious eavesdropper. They use terrestrial base stations as beamformers and friendly jammers utilizing inter-segment interference to reduce the quality of the eavesdropper channel in a 3D wiretap environment. The proposed beamforming schema depends on spatial correlation with energy consumption and secrecy rate constraints to maximize secrecy energy efficiency. On the other hand, authors in \cite{lei2021joint} assume satellite-UAV cognitive network architecture where each network has its users, and they share the spectrum divided into different channels, each channel serving the satellite user and UAV user. They consider terrestrial eavesdroppers wiretap the signals from UAVs and aim to maximize the secrecy rate of UAV users by optimizing power, resource allocation and interference constraints. They compare their scheme with the other two power allocation schemes and show the superiority of the proposed scheme in providing a higher secrecy rate.

\cite{guo2020physical} proposed a user scheduling scheme for multiuser satellite and wireless sensors network to enhance secrecy performance. They drive the closed-form expression of secrecy outage probability and average secrecy capacity when the CSI of the eavesdroppers is unknown. They assume a single antenna for all users and eavesdroppers and utilize the time division multiple access to ensure one user is using the service at each time. Similarly, authors in \cite{yin2021uav} consider multi-beam satellite that enables vehicle communications with UAVs as amplify and forward relays to enhance the main channel between satellite and vehicles where rain attenuation is assumed for the satellite channel. Moreover, UAVs act as a jammer to produce artificial noise that affects the eavesdropper channel. They jointly optimize the power for satellite beamforming and power allocation for UAV relays to maximize the secrecy rate of satellite-to-vehicle links without affecting the QoS requirements. They optimize the non-convex optimization problem using several methods and compare the secrecy rate for the proposed technique with a case without artificial noise and a benchmark case without UAV cooperation to show the improvement in security performance resulting from UAV assistance. 

Table \ref{tab:tableIoT} summarized the discussed works related to PLS on satellite-based IoT networks, which mainly focus on system confidentiality with a single eavesdropper and consider the security of downlinks only. Authenticating satellite and IoT devices using PLS to prevent spoofing and impersonation attack is not studied yet. Moreover, protection against multiple malicious IoT devices and denial of service attacks (which are common in the case of IoT devices) are not available in the literature. Furthermore, more attention should be paid to studying the security of the uplink communication, which is critical for IoT systems since IoT devices transfer a massive amount of data to the satellite, and this data can be eavesdropped or altered if no proper uplink security mechanism is applied. 

\begin{table*}[!t]
\caption{Comparison between existing PLS works for IoT networks  \label{tab:tableIoT}}
\centering
\begin{tabular}{|m{0.7cm}|m{1.6cm}|m{5em}|m{1.2cm}|m{6em}|m{2.2cm}|m{2.4cm}|m{3cm}|}
\hline
Work &  Security Goal &  Network & link Type & Adversary Model & PLS Techniques & PLS feature used  & Security Measurement \\
\hline

\cite{yin2019secrecy} (2019) & Confidentiality& Multi-user Satellite & Downlink & Single eavesdropper & Information theory  & Inter-user interference  & Secrecy rate\\
\hline
\cite{bas2019physical} (2019) & Confidentiality & & Uplink \& Downlink  
 & Single eavesdropper & Information theory  & Time pulses & Secrecy Capacity\\
\hline

\cite{ruan2020cooperative} (2020) & Confidentiality & Cognitive satellite terrestrial network & Downlink  
 & Malicious UAV  & Beamforming  & Spatial correlations & Secrecy energy efficiency 
\\
\hline
\cite{guo2020physical} (2020) & Confidentiality & Multi-user satellite& Downlink 
 & Colluded Passive eavesdropper & Information theory &Time division multiple access   
 &  \begin{tabular}{@{}c@{}} Secrecy Outage Probability \\ Average secrecy capacity\end{tabular}
\\
\hline
\cite{lei2021joint} (2021) & Confidentiality & Cognitive satellite and UAV
& Downlink  
 & Terrestrial eavesdropper & Information theory  &Large-scale CSI
 &  Secrecy rate 
\\
\hline
\cite{yin2021uav} (2022) & Confidentiality & Multi-beam Satellite & Downlink&Terrestrial eavesdropper for each beam &Relaying
\& Artificial noise & Perfect CSI
 &  Secrecy rate 
\\
\hline

\end{tabular}
\end{table*}

\subsection{Domain 3: Navigation Satellite Systems }
\label{GNSS}
Global Navigation Satellite Systems (GNSS) are localization systems that use satellite signals to share information about location, speed, and time with earth users. GPS and GALILEO are examples of GNSS widely used in several applications, such as intelligently connected vehicle networks, mobile applications, and maritime. However, GNSS is vulnerable to spoofing attacks due to its signal's lack of authentication mechanism. Specifically, the publicly known spreading codes allow attackers to forge messages and send them to users as legitimate messages, causing wrong positioning, which can lead to several consequences\cite{papadimitratos2008protection}. The attacker can jam legitimate messages to prevent their reception by legitimate receivers and give more chances to illegitimate ones. Through the years, several mechanisms have been proposed to prevent GNSS spoofing attacks. In this section, we focus on physical layer schemes used to detect and mitigate GNSS spoofing.

Due to the invisibility of changing the algorithms of GNSS, the proposed methods are superposition based. In \cite{formaggio2018authentication}, the proposed scheme uses an authentication message and artificial noise on top of the GNSS message to authenticate it. In this paper, they consider three channels in the system: 1) the navigation channel, which is the broadcast signal from the satellite to all users, 2) the attack channel where the eavesdropper transmits the spoofing signal to the legitimate user, 3) the authenticated channel where a ground segment can communicate with all users through infinite bandwidth and eavesdropper has no control on this channel (authenticated through higher layer protocol). The legitimate user uses the authenticated channel to know the signature and artificial noise after revealing it by the system and check if the received navigation signal coordinates are included. They assume that the attacker receives the signal from the authenticated channel but does not have control over it. Two attack models were considered in the analysis: the generation attack, where the attacker generates a simple navigation signal and uses it, and the replay attack, where the adversary combines the signal with a previously used authentication signal. Although this method utilizes PLS, it depends on the upper layer algorithms to secure the authenticated channel, which is essential for the system design. On the other hand, authors in \cite{jiang2018satellite} utilize satellite signals' radio frequency spatial characteristics to identify spoofing signals. Using the K-means cluster algorithm, they use consultation figures to extract radio features and fingerprint satellite signals. Specifically, the clustering algorithm generates multiple cluster centers according to the signal features, which help detect spoofing by analyzing the center change. They verify the validity of the proposed scheme using experimental analysis through real satellite signals and the spoofed signals generated by the Universal Software Radio Platform.

Similarly, the authors in \cite{foruhandeh2020spotr} use device fingerprinting to create signatures for GPS satellite transmissions and allow identifying the spoofed and authentic signals. The attack model considered in this paper is similar to \cite{formaggio2018authentication}, where spoofer attackers can generate genuine GPS messages or use relay messages to spoof legitimate receivers and induce wrong position, velocity, and time messages. They propose a feature extraction method for fingerprinting based on the GPS receivers' internal hardware architecture. Specifically, they use early‐late phase (ELP) correlator outputs, code-lock-detector (CN0), carrier-lock-detector (CLT) for feature extraction, and Multivariate Normal Distribution (MVN) as a scoring metric to measure similarity. To find thresholds for genuine GPS signals, they first train the feature extraction method using a sample spoofed-free dataset and then test it on spoofed signals to calculate the MVN for each observation, and low MVN means spoofed signal. These thresholds are used offline for real-time spoofing detection after the cross-validation phase. They collect live datasets and perform experimental analysis for the proposed algorithm under different attack scenarios, including location, time, and multipath.

Depending on feature analysis, the authors in \cite{zhu2022global} propose a spoofing detection method using a support vector machine (SVM) to analyze signal quality monitoring (SQM), moving variance (MV), improved SQM moving average (MA), early‐late phase, carrier-to-noise ratio–MV and clock offset rate. After feature extraction, the data is preprocessed using bais and normalization techniques and divided into train and test datasets. They perform offline training and testing phases for SVM using Texas Spoofing Test Battery (TEXBAT) dataset under different Kernal and activation functions to get the best performance. They reach 92.31\% accuracy compared to other detection methods and provide other indices evaluation such as recall, precision, and F1. 

Another detection approach is proposed in \cite{oligeri2020gnss}, where they utilize the Iridium satellite constellation to detect GNSS spoofing using Iridium unencrypted IRA messages that can be received worldwide from all Iridium satellites. In particular, using thousands of messages from the data collection campaign, they reverse engineer satellite parameters such as speed, packet interval time, and satellite coverage. They assume a system model suitable for remote areas such as deserts and oceans where the receiver can not check their location with other sources. Detecting spoofing depends on collecting IRA messages at the receiver side and compensating for the movement. Then the receiver estimates the location using the longitude and latitude of the collected beams. Finally, the estimated position is compared to the position received by the GNSS satellite, and if it is greater than the specific threshold, the signal is considered malicious.

\cite{laverty2021gnss} consider another application scenario where they propose a spoofing detection method for electric Substations since it relays on GNSS services for time transfer. The method utilizes multiple antenna receivers for GPS signals in proximity where the spoofer cannot send a spoofing signal for each antenna, so only a limited number of the antennas will give an incorrect location. Each antenna is connected to a GNSS clock used in a conventional substation that estimates the position using the received signal and then sends it to the detector. The detector will compare all positions received, decide the expected location, and raise the alarm if a spoofing attack is detected. They validate that their method can detect the spoofing attack in 4 to 10 seconds which eliminates the attack effect on the substation. 

Table \ref{tab:tableGNSS} summarizes the discussed works related to PLS in navigation satellite system, which clearly shows the focus on PLS authentication and spoofing detection mechanisms since navigation satellite systems are vulnerable to spoofing and impersonation attacks. The confidentiality of these systems is not considered critical since the messages' are not secret (they only provide location and time information). However, the integrity of these messages should be preserved to avoid malicious alteration during transmission. Moreover, navigation satellite services are vital to many applications; hence more attention should be paid to study the system availability. 

\begin{table*}[!t]
\caption{Comparison between existing PLS works for Navigation Satellite System  \label{tab:tableGNSS}}
\centering
\begin{tabular}{|m{0.7cm}|m{2cm}|m{0.9cm}|m{3cm}|m{1.9cm}|m{2.9cm}|m{3cm}|}
\hline
Work & Security Goal &Satellite Type & Adversary Model & PLS Techniques & PLS feature used  & Security Measurement \\
\hline

\cite{formaggio2018authentication} (2018) & Authentication& Galilo & Generation attack -
Replay attack& Artificial Noise & Random generated codes &  Channel gain 
- Channel estimation error\\
\hline

\cite{jiang2018satellite} (2018) & Spoofing detection& GPS & -& Hardware fingerprinting &  Spatial distribution characteristics & Euclidean distance \\
\hline

\cite{foruhandeh2020spotr} (2020) & Spoofing detection& GPS & Generation attack
- Replay attack& Hardware fingerprinting & \begin{tabular}{@{}c@{}c@{}} Early‐late phase\\ code-lock-detector\\ carrier-lock-detector\end{tabular} & Equal error rates
\\
\hline

\cite{oligeri2020gnss} (2020) & Spoofing detection& GNSS & Generate Fake 
messages using SDR& Parameters Reverse engineering   
 & IRA Iridium messages & False positive rate\\
\hline

\cite{zhu2022global} (2021) & Spoofing detection& GPS & Generate spoofing signal using user tracking signal& Feature extraction
 & \begin{tabular}{@{}c@{}c@{}c@{}c@{}c@{}c@{}} Quality Monitoring (SQM)\\ moving variance (MV)\\ Improved SQM \\Moving average (MA)\\ Early‐late phase\\Carrier-to-noise ratio\\MV and clock offset rate\end{tabular}  & Detection accuracy\\
\hline

\cite{laverty2021gnss} (2022) & Spoofing detection& GNSS &Generate spoofing signals for electric substation& Multiple antenna
 & spatial area & Detection speed\\
\hline
\end{tabular}
\end{table*}

\subsection{Domain 4: Free Space Optical Satellite Communication}
\label{FSO}
Optical Wireless Communication (OWC) uses optical carriers such as visible light and Infrared to transmit information. Free Space Optical (FSO) is a type of OWC that uses directive laser beams to transmit data wirelessly in unguided free space media to provide high-speed communication between two points. The Optical medium is an unlicensed spectrum that overcomes the limitations and congestion of the current Radio frequency band and suffers from less electromagnetic interference \cite{jahid2022contemporary}. Furthermore, FSO can provide a high data rate with low latency for several kilometers without expensive infrastructure, reducing installation and maintenance costs. Satellite communications, including uplink, downlink, and interlinks, start incorporating FSO links to leverage its advanced features and improvements. Specifically, FSO allows higher data rates to reach Gigabits per second (Gbps), significantly better compared to the RF band. Moreover, FSO  has inherent security features that it can not penetrate walls and can support quantum cryptography when fiber optic infrastructure is unavailable \cite{alshaer2021reliability}.  
 
On the other hand, FSO suffers from fading and attenuation induced by atmospheric turbulence such as fog, rain, and snow \cite{khalighi2014survey}. In addition, beam divergence caused by beam wanders causes misalignment between the transmitter and the receiver. These conditions can affect the link's reliability and availability and limit its capacity to short-distance ranges. To tackle these shortcomings in FSO and leverage its features simultaneously, a hybrid Satcom system between RF and FSO is designed. RF links are more robust towards weather conditions and mature than FSO development. Several hybrid Satcom designs are proposed in the literature, such as hard switching \cite{swaminathan2021haps}, adaptive switching \cite{shah2021adaptive}, and HAPS relay-based design. In this section, we discuss the physical layer security analysis available in the literature for FSO and hybrid FSO RF systems.    

Authors in \cite{illi2020phy}\cite{hayashi2020physical} design an FSO satellite system with ground optical stations. In \cite{illi2020phy}, the ground stations send $N$ users information by optical feeder link to the satellite, which acts as a relay that decodes the optical wave received and sends it as beams to multiple earth users with earth eavesdropper for both links. However, authors in \cite{hayashi2020physical} consider the uplink connection between the ground station and a GEO satellite with a nano-satellite eavesdropper. They propose two PLS secrecy coding protocols: one-way and two-way protocols with two types of photodetectors: threshold and PNR. Their two-way scheme ensures secrecy even when the eavesdropper's channel is better than the legitimate channel. Both works analyze the secrecy capacity using the SNR between the legitimate user and eavesdropper. In \cite{illi2020phy}, the authors consider two scenarios, ZF coding and without ZF coding, for the analysis. Without ZF coding, the satellite decodes the data received and directly forwards it to the users, while with ZF coding, the satellite code the beams using the ZF technique before sending. The results showed that the ZF technique improves the security capacity with specific positions.

In \cite{ai2019physical}, the authors consider a hybrid system RF link between the satellite and the ground station (the long distance) and an FSO link between the station and destination (short distance). The station is the decode and forward relay between source and destination, and the eavesdropper is assumed to be on the earth trying to get access to the data sent by satellite. They analyze the Average Secrecy Capacity and Secrecy Outage Probability under different satellite and FSO channel conditions, relaying schemes, and FSO detection methods. RF link conditions have more impact on the system secrecy than the FSO, while secrecy diversity depends on FSO conditions when using amplify and forward relaying methods.  

Both \cite{yahia2021use} and \cite{odeyemi2022mixed} use a high-altitude platform station (HAPS) as a relay between an LEO satellite and the earth's destination. The HAPS has an FSO downlink with the satellite and an RF downlink with the earth station. The eavesdropper in this system is passive and present on the earth's surface. They drive the secrecy outage probability (SOP) and the probability of positive secrecy capacity of the system and analyze them under different channel conditions. The results in \cite{yahia2021use} showed that the RF link has more impact on the system performance. At the same time, both analyses emphasize the criticality of pointing errors and channel shadowing on the system's security. 

Table\ref{tab:tableFSO} summarizes the works related to PLS on FSO satellite communication, which shows that the use of PLS in optical satellite communication is still limited as most works consider a hybrid system of optical and radio frequencies and relay techniques between them. Therefore, the PLS technique needs to be investigated more in FSO links with a foucs on how distance and weather conditions can affect it. 

\begin{table*}[!t]
\caption{Comparison between existing PLS works for FSO satellite communication  \label{tab:tableFSO}}
\centering
\begin{tabular}{|m{0.7cm}|m{1.6cm}|m{1.2cm}|m{2.8cm}|m{3.1cm}|m{4.4cm}|}

\hline
Work & Security Goal & Link Type  & Relaying Scheme 
&Channels Types& Security Measurement \\
\hline
\cite{ai2019physical} (2019) & Confidentiality & Downlink  & \begin{tabular}{@{}c@{}}Amplify and forward
 \\ Decode and forward \end{tabular} & \begin{tabular}{@{}c@{}} Shadowed-Rician For RF 
 \\ GammaGamma for optical \end{tabular} &  \begin{tabular}{@{}c@{}}Average Secrecy Capacity
 \\ Secrecy Outage Probability\end{tabular}
 \\
\hline
\cite{illi2020phy} (2020) & Confidentiality & Downlink and uplink
 & Decode and forward &  GammaGamma for optical &  Secrecy Capacity
 \\
 \hline
 \cite{hayashi2020physical} (2020) & Confidentiality & Uplink  & - & Poisson channel & Secrecy Capacity 
 \\
\hline
\cite{yahia2021use} (2021) & Confidentiality & Downlink  & Decode and forward & \begin{tabular}{@{}c@{}} Shadowed-Rician For RF 
 \\ GammaGamma for optical \end{tabular} &  \begin{tabular}{@{}c@{}}Secrecy outage probability 
 \\ Probability of positive secrecy capacity\end{tabular}
 \\
\hline

\cite{odeyemi2022mixed} (2022) & Confidentiality & Downlink  & Amplify and forward & \begin{tabular}{@{}c@{}} Shadowed-Rician For RF 
 \\ GammaGamma for optical \end{tabular} &  \begin{tabular}{@{}c@{}}Secrecy outage probability 
 \\ Probability of positive secrecy capacity\end{tabular}
 \\
\hline
\end{tabular}
\end{table*}

\subsection{Domain 5: Inter-satellite Communication }
\label{inter}
Inter-satellite communication links connect satellites to each other. Typically, satellites are deployed in the formation of many satellites where communication between them is essential for service continuity. Satellites must share their navigational and mobility data to keep the formation in the correct movement \cite{radhakrishnan2016survey}. The connection between satellites increases network throughput, where the message is routed through satellites until it reaches its destination without depending on ground stations. Specifically, this communication creates a network in the space which improves the transmission and supports the service handover. On the other hand, Inter-satellite links are complex since they connect satellites from various levels (GEO, MEO, and LEO), with different speeds and altitudes that can affect the link availability and quality. These links can be based on radio frequency or optical communication.

The complexity of inter-satellite links increases its exposure to adversaries from multiple levels, such as satellites from other constellations, spacecraft, and adversaries on earth. In addition, inter-satellite links have different channel states than ground links because of the high mobility of the satellites and fewer fading sources. Despite this complexity and the implications of the inter-satellite links, security mechanisms to protect the space networks are still limited. Inter-satellite links are vulnerable to attacks against confidentiality and authentication where the identity of satellites is not identified before initiating a connection, and data are not protected against any adversary in middle communication. Furthermore, adversaries can target the availability of inter-satellite links to cause service interruption by overwhelming the link using a compromised satellite or malicious spacecraft.

Some authors apply the PLS framework to inter-satellite links such as \cite{topal2021securing} \cite{topal2022physical}. In \cite{topal2021securing}, the authors use Doppler frequency shift to generate keys for securing the inter-spacecraft links. While in \cite{topal2022physical} they use the Doppler frequency shift as a physical layer authentication method for satellite identification, utilizing satellite voting. Doppler frequency shift is the change of frequency relative to the motion. Both works use radio frequency links and satellites' mobility information, such as speed and locations, to estimate the Doppler frequency shift of each satellite. Since the mobility information of each satellite is already shared on the network, this method does not introduce overhead or additional channels.

\begin{table*}[!t]
\caption{Comparison between existing PLS works for Inter-satellite communication 
\label{tab:tableInter}}
\centering
\begin{tabular}{|c|c|c|c|c|}
\hline
Work & Link Type & Security Goal & PLS Feature Used  & Security Measurement \\
\hline
\cite{topal2021securing} (2021) & \begin{tabular}{@{}c@{}}Radio frequency \\ inter-spacecraft link\end{tabular} & Secret Key generation & Doppler frequency shift 
 & \begin{tabular}{@{}c@{}}Maximum Achievable Key Rate \\ Key Disagreement rate link\end{tabular}
 \\
\hline
\cite{topal2022physical} (2022) & \begin{tabular}{@{}c@{}}Radio frequency \\ inter-satellite link\end{tabular} & Authentication & Doppler frequency shift & \begin{tabular}{@{}c@{}}Spoofing Detection Rate \\ False Alarm Probability\end{tabular}
 \\
\hline
\end{tabular}
\end{table*}

In \cite{topal2021securing}, the authors assume communication between two spacecrafts, a transmitter and a receiver, in deep space using a time division duplexing (TDD) system. Both legitimate users will experience identical Doppler frequency shifts using their high mobility. At the same time, the eavesdropper observes a different shift since her relative velocity and position vary from the transmitter and receiver, which allows them to use the Doppler frequency shift as a source of secret keys. The secret keys generation process starts with pilot transmission and reception between the transmitter and receiver, where it is assumed they have the same velocity at this step. Then each spacecraft will estimate the nominal power spectral density for each communication block separately, which measures the power of the content of the message versus its frequency. After that, they will quantize the achieved value to get the raw secret sequence. The authors perform a Monte-Carlo simulation for the proposed scheme and analysis several aspects, such as the Key Disagreement rate and Maximum Achievable Key Rate, to prove the practicality and robustness of the technique. 

In \cite{topal2022physical}, the authors consider a set of legitimate LEO satellites and one satellite that needs to be authenticated. When satellites receive a message from the suspect satellite, they will calculate the Doppler frequency using nominal power spectral density estimation, which measures the power of the content of the message versus its frequency. All satellites will compare the resulting Doppler frequency with the expected one for the transmitter, which can be calculated by any satellite since the transmitter's location, and speed are known to them. Then each satellite will decide if this satellite is a transmitter or an eavesdropper according to this comparison and specific threshold and send this decision to the fusion center. In the fusion center, the algorithm will collect all decisions and use the decision method to make the final judgment on the satellite's identity. The authors analyze algorithm performance with different numbers of satellites and lengths for the time slot. They also compare three fusion decision methods: OR, AND, and Majority rule, and measure their spoofing detection and false alarm probabilities. The results showed that the majority rule outperforms others, where it achieved a high spoofing detection rate with low false alarm probability.

PLS in optical inter-satellite links is still an uncovered area in the literature due to its intrinsic security features earned by the inability to penetrate through walls. However, optical links are yet vulnerable to attacks such as eavesdropping as authors in \cite{yahia2022optical} propose eavesdropping techniques for space networks where HAPS is the eavesdropper for LEO satellites communication.

Table \ref{tab:tableInter} summarizes recent work related to PLS for inter-satellite communication. Although inter-satellite communication is critical to the satellite system, its security is not yet well studied in the literature. Both radio and optical inter-satellite links are vulnerable to multiple attacks, such as eavesdropping, spoofing, and jamming. 

In LEO satellite constellations, the situation is exacerbated, where attacking these inter-satellite links will often affect major services, especially those requiring the transfer of satellite-to-satellite messages. 

Moreover, inter-satellite links exhibit unique features that differ from satellite-to-earth links, which prevent the usage of the available PLS techniques in the satellite-to-earth setting. Hence, more investigation is required to study how these features can be leveraged using PLS to provide security against possible attacks.

\section{Research gaps and future directions}
\label{gaps}

Physical layer security in satellite systems still has many open issues that need to be investigated. In this section, we highlight open research opportunities and future directions that we identify through our survey study.

\subsection{Availability schemes}
Due to the importance of satellite systems in the next emerging network architecture, ensuring their availability is essential to avoid system outages or interruptions. Satellite systems can suffer denial of service attacks if the satellite or link is overwhelmed with malicious packets that the satellite cannot handle and cause service outages for legitimate users. Moreover, anti-jamming techniques for satellite communications need to be investigated since it is a wireless signal that adversaries can target through multiple types of jamming, causing transformations to the signals that prevent correct decoding and processing. PLS techniques are promising in anti-jamming methods and cooperative jamming to prevent various attacks in 6G networks \cite{vaishnavi2021survey}. Therefore, more works should consider PLS availability techniques in satellite scenarios to study its feasibility with the unique physical characteristics of satellite communication. 

\subsection{Uplink secrecy techniques}
The current literature mainly focuses on securing the downlink communication from the satellite to other receivers. However, uplink communication from any terrestrial or aerial device to the satellite is vulnerable to several attacks such as spoofing and alteration by malicious adversaries. Especially in  
broadband satellite communication, the uplink handles a huge amount of data sent by users to the satellite, and this data must be secured through schemes suitable for heterogeneous devices and scenarios. Moreover, the uplink communication for the satellite can be used to overwhelm the satellite using multiple techniques of denial of service attack. Hence, researchers should give more attention to securing uplink communication using PLS, considering performance and power constraints. 
 
 \subsection{Smart threat models}
Most current literature assumes that the eavesdropper is passive with limited resources and a single antenna and does not actively affect the main channel. They do not consider the case where eavesdroppers can extract CSI of the main channel by active interception, reducing the current schemes' effectiveness. These assumptions are becoming ideal with advancements in adversaries' resources, such as the number of antenna and signal processing techniques. Moreover, the widespread assumption of static and not moving eavesdroppers can be enhanced where we can have dynamic eavesdroppers with unknown locations such as UAVs. Hence, more complex adversary models should be considered in satellite systems to cooperate with advancing eavesdropping capabilities and design more robust solutions.

\subsection{Authentication and Anti-spoofing techniques}
Through analysis of the available schemes for authenticating satellite signals, we notice that most techniques are focused on GNSS satellites due to their significance and vulnerability to spoofing. However, other types of satellites used in other services, such as broadband and IoT connection, are vulnerable to spoofing or impersonation attacks that affect their service and give access to unauthorized devices. In particular, failing to authenticate the messages sent to the satellite can cause several consequences due to processing malicious messages/ code received. In wireless systems such as satellites, the adversary sends signals with higher power than the legitimate user to deceive the receiver into considering his spoofed message as the legitimate one\cite{yilmaz2015survey}. Hence, the adversary will get access to the service and information intended for the legitimate user. As a result, anti-spoofing PLS strategies need investigation in satellite scenarios with different services to prevent unauthenticated access or information leakage.

\subsection{Integrity techniques }
Message manipulation detection techniques using PLS are not considered in the literature yet for any wireless communication. Integrity violation attacks such as man-in-the-middle intercept the message during transmission, modify it or manipulate it, and then re-transmit it. This change in the message needs to be detected by the receiver to prevent wrong information transmission and processing. Satellite systems are vulnerable to these attacks where adversaries can intercept the signal sent by satellite to the terrestrial user or vice-versa and manipulate it either to disturb the service or to get access by exploiting these messages. PLS techniques to achieve integrity needs to be investigated using different wireless signals, including satellite systems. 

\subsection{PLS for Inter-satellite link communication }
Inter-satellite communication is essential for providing high-speed communication and increasing network throughput by connecting satellites to each other. With the increasing interest in LEO satellites that are relatively near to the earth and advancements in adversary power, attacks on inter-satellite communication have become feasible. There is limited focus on securing inter-satellite communications in the literature for both RF and optical links. In particular, inter-satellite links have unique physical layer characteristics in terms of fading, doppler shift, and weather condition effects that can be exploited to secure communication. Moreover, optical inter-satellite links provide high speed and bandwidth, but it is still vulnerable to attacks such as eavesdropping, and there are no proposed solutions yet in the literature. PLS strategies are promising techniques that need to be studied extensively in inter-satellite communication scenarios by investigating their physical features and possible attacks. 

\subsection{Machine learning based PLS}
Wireless networks are becoming more complex with heterogeneous devices and high mobility requirements. Recently, machine learning techniques have emerged to support PLS in handling network complexity by intelligently handling signal processing and feature exploitation. Several supervised learning and unsupervised ML techniques are used for intelligent PLS in different categories, such as physical layer authentication, antenna selection, and relay node selection\cite{kamboj2021machine}. Proposed ML-based PLS schemes consider various wireless communications, including device-to-device, cognitive networks, and Non-orthogonal multiple access (NOMA). However, ML-based algorithms are not explored in satellite systems scenarios, and there is no analysis of the applicability of the proposed methods in the satellite case. Therefore more investigation is required to benefit from ML techniques in simplifying PLS in satellite systems and consider its unique features. 

\section{Conclusion}
Satellite communication systems are vulnerable to various attacks due to their open nature and the heterogeneity of the connected devices. PLS schemes are lightweight security approaches that exploit physical layer characteristics to provide the required security while ensuring minimal overhead, making them suitable for several communication types, including satellite systems. 

In this paper, we first classify modern satellite communication networks into five domains and then analyze the use of PLS in each domain; these domains are:  hybrid satellite-terrestrial networks, satellite-based IoT, navigation satellite systems, FSO-based satellites, and Inter-satellite communications. After providing the necessary background about satellite networks and PLS techniques, we review the state-of-the-art of PLS solutions in each domain and compare their approaches, adversary models, and results. Finally, we highlight a few open research gaps for PLS in satellite systems and propose a few potential future directions, which we believe the community must focus on.

\section*{Acknowledgments}
This publication was made possible by GSRA grant \# GSRA8-L-2-0521-21046 from the Qatar National Research Fund (a member of Qatar Foundation) and a TÜBİTAK-QNRF Joint funded grant \# AICC03-0530-200033. The findings herein reflect the work, and are solely the responsibility, of the authors.

This work has been submitted to the IEEE for possible publication. Copyright may be transferred without notice, after which this version may no longer be accessible.

\printbibliography

\begin{IEEEbiographynophoto} 
{\bf Nora Abdelsalam } received a Bachelor of Science in Computer Engineering from Qatar University, Qatar in 2019.  She is currently working on her Masters degree in Cyber Security at Hamad Bin Khalifa University. Her research interests include Physical Layer Security, Satellite Communications and Machine Learning.
\end{IEEEbiographynophoto}

\begin{IEEEbiography} [{\includegraphics[width=1in,height=1.25in,clip,keepaspectratio]{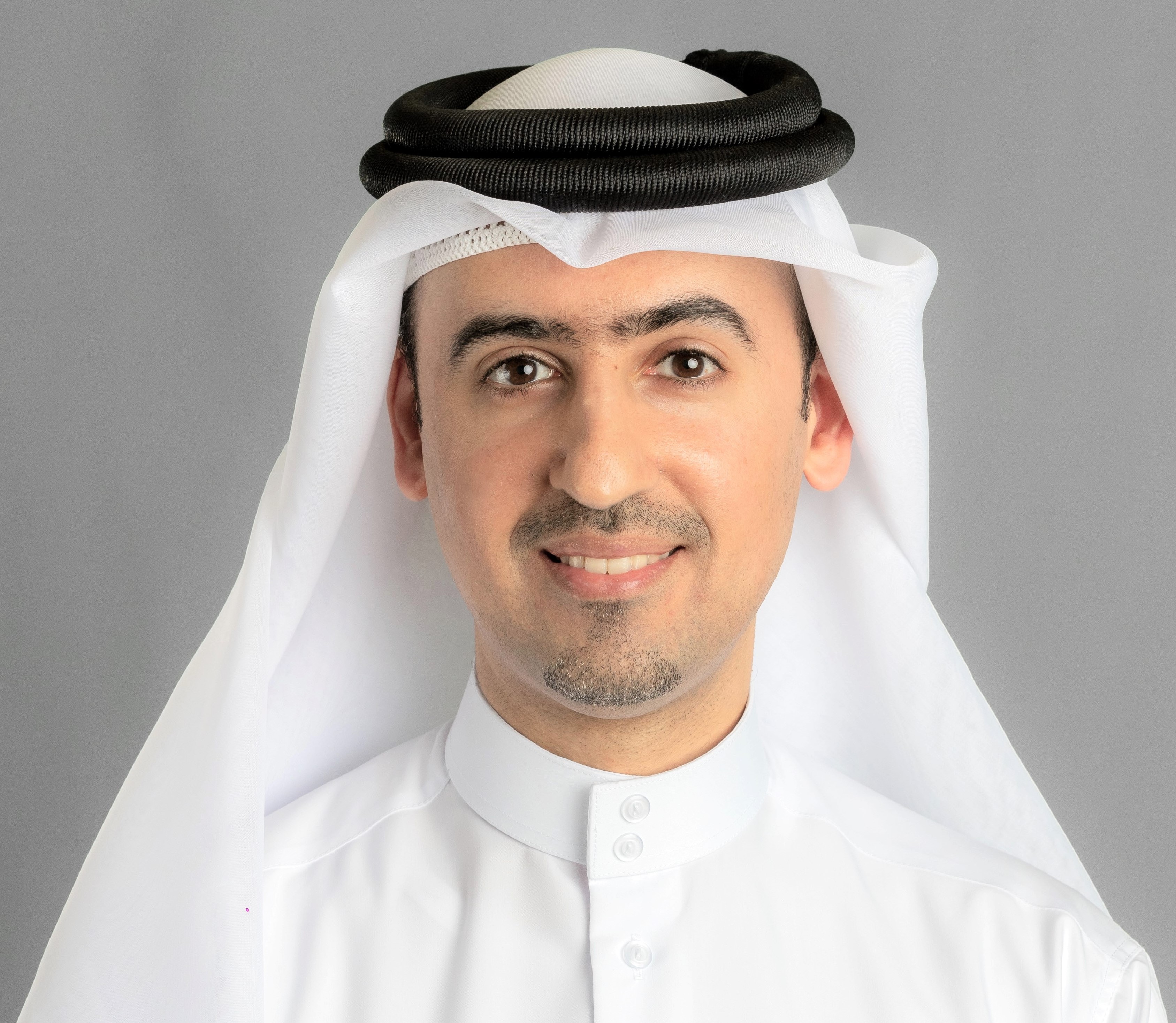}}]
{\bf Saif Al-Kuwari} received a Bachelor of Engineering in Computers and Networks from the University of Essex, UK in 2006, a two PhD's from the University of Bath and Royal Holloway, the University of London in Computer Science, both in 2011. He is currently an Assistant Professor at the College of Science and Engineering at Hamad Bin Khalifa University.  His research interests include Applied Cryptography, Quantum Computing, Computational Forensics, and their connections with Machine Learning. He is IET and BCS fellow, and IEEE and ACM senior member.
\end{IEEEbiography}

\begin{IEEEbiography} [{\includegraphics[width=1in,height=1.25in,clip,keepaspectratio]{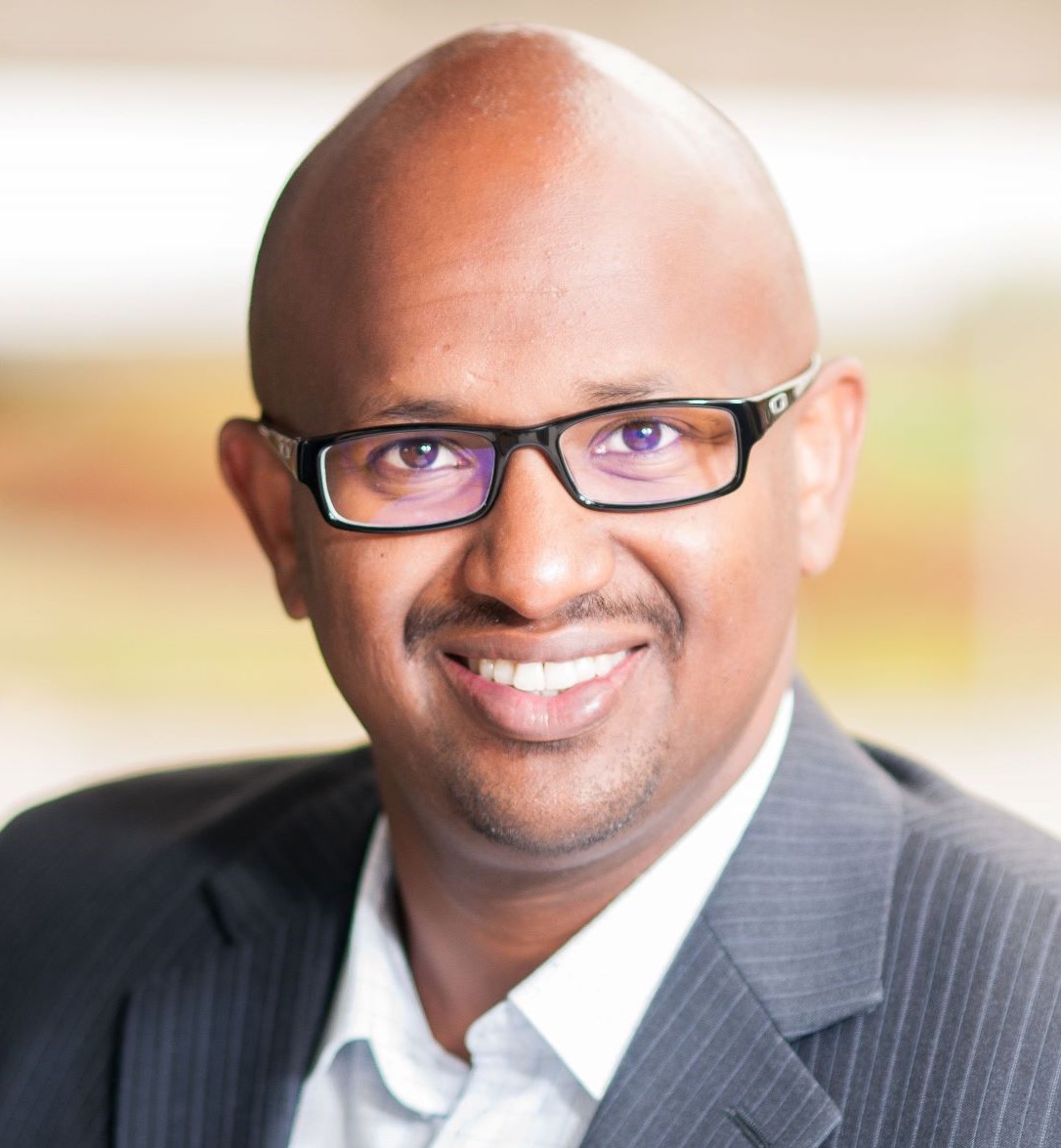}}]
{\bf Aiman Erbad} is an Associate Professor in the College of Science and Engineering, Hamad Bin Khalifa University (HBKU). He obtained a Ph.D. in Computer Science from the University of British Columbia (Canada), and a Master of Computer Science in embedded systems and robotics from the University of Essex (UK). Aiman also received the 2020 Best Research Paper Award from Computer Communications, the IWCMC 2019 Best Paper Award, and the IEEE CCWC 2017 Best Paper Award. His research interests span cloud computing, edge intelligence, Internet of Things, secure networks, and distributed systems. He is a senior member of IEEE and ACM.
\end{IEEEbiography}

\vfill

\end{document}